\documentclass[trackchanges,twocolumn,twocolappendix]{aastex631}

\usepackage{gensymb}
\usepackage{amsmath}

\newcommand{\vsys}{v_{\text{sys}}}
\newcommand{\kms}{km~s$^{-1}$}
\newcommand{\mjbeam}{mJy~beam$^{-1}$}
\newcommand{\mjbeamkms}{mJy~beam$^{-1}$~km~s$^{-1}$}

\received{June 25, 2023}
\revised{September 6, 2023}
\accepted{September 6, 2023}

\shorttitle{Oph IRS63}
\shortauthors{Flores et al.}

\begin{document}

\title{Early Planet Formation in Embedded Disks (eDisk) XII:\\ Accretion streamers, protoplanetary disk, and outflow in the Class I source Oph IRS63}

\email{caflores@asiaa.sinica.edu.tw}
\correspondingauthor{Christian Flores}
\author[0000-0002-8591-472X]{Christian Flores}
\author[0000-0003-0998-5064]{Nagayoshi Ohashi}
\affiliation{Academia Sinica Institute of Astronomy \& Astrophysics, 11F of Astronomy-Mathematics Building, AS/NTU, No.1, Sec. 4, Roosevelt Rd,Taipei 10617, Taiwan, R.O.C.}

\author[0000-0002-6195-0152]{John J. Tobin}
\affil{National Radio Astronomy Observatory, 520 Edgemont Rd, Charlottesville, VA, 22903, USA} 

\author[0000-0001-9133-8047]{Jes K. J\o rgensen}
\affiliation{Niels Bohr Institute, University of Copenhagen, \O ster Voldgade 5-7, 1350, Copenhagen, Denmark}

\author[0000-0003-0845-128X]{Shigehisa Takakuwa}
\affiliation{Department of Physics and Astronomy, Graduate School of Science and Engineering, Kagoshima University, 1-21-35 Korimoto, Kagoshima,Kagoshima 890-0065, Japan}
\affiliation{Academia Sinica Institute of Astronomy \& Astrophysics, 11F of Astronomy-Mathematics Building, AS/NTU, No.1, Sec. 4, Roosevelt Rd,Taipei 10617, Taiwan, R.O.C.}

\author[0000-0002-7402-6487]{Zhi-Yun Li}
\affiliation{University of Virginia, 530 McCormick Rd., Charlottesville, Virginia 22904, USA}

\author[0000-0001-7233-4171]{Zhe-Yu Daniel Lin}
\affiliation{University of Virginia, 530 McCormick Rd., Charlottesville, Virginia 22904, USA}

\author[0000-0002-2555-9869]{Merel L.R. van 't Hoff}
\affil{Department of Astronomy, University of Michigan, 1085 S. University Ave., Ann Arbor, MI 48109-1107, USA}

\author[0000-0002-9912-5705]{Adele L.\ Plunkett}
\affiliation{National Radio Astronomy Observatory, 520 Edgemont Rd, Charlottesville, VA, 22903, USA}

\author[0000-0003-4099-6941]{Yoshihide Yamato}
\affiliation{Department of Astronomy, Graduate School of Science, The University of Tokyo, 7-3-1 Hongo, Bunkyo-ku, Tokyo 113-0033, Japan}

\author[0000-0003-4361-5577]{Jinshi Sai (Insa Choi)}
\affiliation{Academia Sinica Institute of Astronomy \& Astrophysics, 11F of Astronomy-Mathematics Building, AS/NTU, No.1, Sec. 4, Roosevelt Rd,Taipei 10617, Taiwan, R.O.C.}

\author[0000-0003-2777-5861]{Patrick M. Koch}
\affiliation{Academia Sinica Institute of Astronomy \& Astrophysics, 11F of Astronomy-Mathematics Building, AS/NTU, No.1, Sec. 4, Roosevelt Rd,Taipei 10617, Taiwan, R.O.C.}

\author[0000-0003-1412-893X]{Hsi-Wei Yen}
\affiliation{Academia Sinica Institute of Astronomy \& Astrophysics, 11F of Astronomy-Mathematics Building, AS/NTU, No.1, Sec. 4, Roosevelt Rd,Taipei 10617, Taiwan, R.O.C.}

\author[0000-0003-3283-6884]{Yuri Aikawa}
\affiliation{Department of Astronomy, Graduate School of Science, The University of Tokyo, 7-3-1 Hongo, Bunkyo-ku, Tokyo 113-0033, Japan}

\author[0000-0002-8238-7709]{Yusuke Aso}
\affiliation{Korea Astronomy and Space Science Institute, 776 Daedeok-daero, Yuseong-gu, Daejeon 34055, Republic of Korea}

\author[0000-0003-4518-407X]{Itziar de Gregorio-Monsalvo}
\affiliation{European Southern Observatory, Alonso de Cordova 3107, Casilla 19, Vitacura, Santiago, Chile}

\author[0000-0002-2902-4239]{Miyu Kido}
\affiliation{Department of Physics and Astronomy, Graduate School of Science and Engineering, Kagoshima University, 1-21-35 Korimoto, Kagoshima,Kagoshima 890-0065, Japan}

\author[0000-0003-4022-4132]{Woojin Kwon}
\affiliation{Department of Earth Science Education, Seoul National University, 1 Gwanak-ro, Gwanak-gu, Seoul 08826, Republic of Korea}
\affiliation{SNU Astronomy Research Center, Seoul National University, 1 Gwanak-ro, Gwanak-gu, Seoul 08826, Republic of Korea}

\author[0000-0003-3119-2087]{Jeong-Eun Lee}
\affiliation{Department of Physics and Astronomy, Seoul National University, 1 Gwanak-ro, Gwanak-gu, Seoul 08826, Korea}

\author[0000-0002-3179-6334]{Chang Won Lee}
\affiliation{Division of Astronomy and Space Science, University of Science and Technology, 217 Gajeong-ro, Yuseong-gu, Daejeon 34113, Republic of Korea}
\affiliation{Korea Astronomy and Space Science Institute, 776 Daedeok-daero, Yuseong-gu, Daejeon 34055, Republic of Korea}

\author[0000-0002-4540-6587]{Leslie W. Looney}
\affiliation{Department of Astronomy, University of Illinois, 1002 West Green St, Urbana, IL 61801, USA}

\author[0000-0001-6267-2820]{Alejandro Santamaría-Miranda}
\affiliation{European Southern Observatory, Alonso de Cordova 3107, Casilla 19, Vitacura, Santiago, Chile}

\author[0000-0002-0549-544X]{Rajeeb Sharma}, 
\affiliation{Niels Bohr Institute, University of Copenhagen, \O ster Voldgade 5-7, 1350, Copenhagen, Denmark}

\author[0000-0003-0334-1583]{Travis J. Thieme}
\affiliation{Institute of Astronomy, National Tsing Hua University, No. 101, Section 2, Kuang-Fu Road, Hsinchu 30013, Taiwan}
\affiliation{Center for Informatics and Computation in Astronomy, National Tsing Hua University, No. 101, Section 2, Kuang-Fu Road, Hsinchu 30013, Taiwan}
\affiliation{Department of Physics, National Tsing Hua University, No. 101, Section 2, Kuang-Fu Road, Hsinchu 30013, Taiwan}

\author[0000-0001-5058-695X]{Jonathan P. Williams}
\affiliation{Institute for Astronomy, University of Hawai‘i at Mānoa, 2680 Woodlawn Dr., Honolulu, HI 96822, USA}

\author[0000-0002-9143-1433]{Ilseung Han}
\affiliation{Division of Astronomy and Space Science, University of Science and Technology, 217 Gajeong-ro, Yuseong-gu, Daejeon 34113, Republic of Korea}
\affiliation{Korea Astronomy and Space Science Institute, 776 Daedeok-daero, Yuseong-gu, Daejeon 34055, Republic of Korea}

\author[0000-0002-0244-6650]{Suchitra Narayanan}
\affiliation{Institute for Astronomy, University of Hawai‘i at Mānoa, 2680 Woodlawn Dr., Honolulu, HI 96822, USA}

\author[0000-0001-5522-486X]{Shih-Ping Lai}
\affiliation{Institute of Astronomy, National Tsing Hua University, No. 101, Section 2, Kuang-Fu Road, Hsinchu 30013, Taiwan}
\affiliation{Center for Informatics and Computation in Astronomy, National Tsing Hua University, No. 101, Section 2, Kuang-Fu Road, Hsinchu 30013, Taiwan}
\affiliation{Department of Physics, National Tsing Hua University, No. 101, Section 2, Kuang-Fu Road, Hsinchu 30013, Taiwan}
\affiliation{Academia Sinica Institute of Astronomy \& Astrophysics, 11F of Astronomy-Mathematics Building, AS/NTU, No.1, Sec. 4, Roosevelt Rd,Taipei 10617, Taiwan, R.O.C.}



\begin{abstract}
We present ALMA observations of the Class I source Oph IRS63 in the context of the Early Planet Formation in Embedded Disks (eDisk) large program. Our ALMA observations of Oph IRS63 show a myriad of protostellar features, such as a shell-like bipolar outflow (in $^{12}$CO), an extended rotating envelope structure (in $^{13}$CO), a streamer connecting the envelope to the disk (in C$^{18}$O), and several small-scale spiral structures seen towards the edge of the dust continuum (in SO). By analyzing the velocity pattern of $^{13}$CO and C$^{18}$O, we measure a protostellar mass of $\rm M_\star = 0.5 \pm 0.2 $~$\rm M_\odot$ and confirm the presence of a disk rotating at almost Keplerian velocity that extends up to $\sim260$ au. These calculations also show that the gaseous disk is about four times larger than the dust disk, which could indicate dust evolution and radial drift. Furthermore, we model the C$^{18}$O streamer and SO spiral structures as features originating from an infalling rotating structure that continuously feeds the young protostellar disk. We compute an envelope-to-disk mass infall rate of $\sim 10^{-6}$~$\rm M_\odot \, yr^{-1}$ and compare it to the disk-to-star mass accretion rate of $\sim 10^{-8}$~$\rm M_\odot \, yr^{-1}$, from which we infer that the protostellar disk is in a mass build-up phase. At the current mass infall rate, we speculate that soon the disk will become too massive to be gravitationally stable.
\end{abstract}



\section{Introduction} \label{sec:intro}

Protoplanetary disks are known to be the birthplace of planets, and therefore their geometrical structure provides key insights into the planet formation process \citep[][and references therein]{Andrews2020}. In the past decade, the radial structure of Class II disks has been thoroughly studied in both gas and dust components thanks in large part to the Atacama Large Millimeter/submillimeter Array (ALMA) \citep[e.g.,][]{Andrews2018,Long2019}. These observations have revealed a prevalent occurrence of sub-structures and asymmetries, such as rings and spiral arms, which in turn, have been interpreted primarily (but not uniquely) as signposts of planet-disk interactions \citep[e.g.,][]{Dong2016, Christiaens2014, Pinte2019, Zhang2015, Gonzalez2017}. If small-scale structures and asymmetries are already present at the age of most Class II disks i.e., from 1 to 5 Myrs, then it is natural to ask when did these features develop? 

To answer this, \cite{Ohashi2023} presented the ALMA Large Program ``Early Planet Formation in Embedded Disks (eDisk)'' that aims to systematically characterize the structures of disks around a sample of 19 nearby ($d<200$~pc) protostars in Class~0/I stages at 1.3 mm at an average spatial resolution of 7~au.

IRAS 16285-2355, also known as Oph IRS63, is a Class~I source of the eDisk sample, and is located in an isolated compact region in the L1709 sub-cloud in the Ophiuchus star-forming region \citep{Ridge2006}. The distance to Oph~IRS63 is estimated to be $132 \pm 6 $~pc based on its proximity to the B45 molecular cloud   \citep[see][galactic coordinates $l=355$, $b=16.1$]{Zucker2020}. The bolometric temperature and luminosity of Oph~IRS63 ($\rm T_{bol}$=348~K and $\rm L_{bol}$=1.3~$\rm L_\odot$), make it the most evolved source (closer to the Class~II phase) of the whole eDisk sample \citep{Ohashi2023}, yet Oph~IRS63 is the youngest source with known multiple rings/gaps in the dust continuum. \cite{Segura-Cox2020} reported high-angular resolution ($0\farcs05 \times 0\farcs03$) ALMA 1.3~mm observations of the disk around Oph IRS63, and they detected low-contrast but detectable annular structures in the dust disk. The ring and gap structures detected by \cite{Segura-Cox2020} make Oph~IRS63 a prime target for the study of early planet formation in embedded disks.



As for molecular lines, \cite{Brinch2013} published sub-millimeter array (SMA) observations of HCO$^+$ J=3–2, with a spectral resolution of  0.23~km s$^{-1}$ and synthesized beam of $\sim1\arcsec$. They used detailed radiative transfer models to estimate the envelope and disk masses of 0.07~$M_\odot$ and 0.099~$M_\odot$, respectively. Furthermore, they modeled the HCO$^+$ observations using a Keplerian disk plus infalling envelope model and derived a stellar mass of $0.8$~$M_\odot$, confirming that Oph~IRS63 is a low-mass protostar.

In this work, we present new ALMA data for the Class~I protostar Oph~IRS63. Our ALMA observations and data reduction are presented in Section \ref{sec:Observations}. Observational results from the continuum and molecular maps are presented in Section \ref{sec:results}. This section is intensively focused on the gas, and we subdivide it into the disk, the envelope, spiral structures, and the outflow. In Section \ref{sec:Analysis}, we propose a  streamer model to reproduce the C$^{18}$O and SO spiral-like features, and we perform a kinematic analysis of the disk and envelope to confirm the presence of a disk rotating at almost Keplerian speed. The discussion is presented in Section \ref{sec:Discussion}, where we focus on the amount of material accumulating in the disk. Finally, we give a brief summary of our findings in Section \ref{sec:Summary}.

\section{Observations and Data Reduction} \label{sec:Observations}

\begin{deluxetable*}{lcccccc}
\tablecaption{Summary of continuum and line observations \label{tab:observation_info}}
\tablecolumns{7}
\tabletypesize{\footnotesize}
\tablehead{
\colhead{Continuum/Line} & \colhead{Frequency} & \colhead{Beam size} & 
\colhead{Velocity resolution} & \colhead{rms} & \colhead{peak intensity} & \colhead{Robust} \\
\colhead{} & \colhead{(GHz)} & \colhead{} & \colhead{(km s$^{-1}$)} & \colhead{(mJy beam$^{-1}$)} & \colhead{(mJy beam$^{-1}$)} & \colhead{}
}
\startdata
1.3 mm continuum & 225 & $0\farcs034 \times 0\farcs025$ ($ 56.7^\circ$) & - & 0.017 & 5.8 & 0\\
$^{12}$CO $J=2$--1 & 230.5380000 & $0\farcs 278 \times 0\farcs 398$ ($78.8^\circ$) & 0.635 & 2.1 & 238 & 0.5 \\
$^{13}$CO $J=2$--1 & 220.3986842 & $0\farcs 291 \times 0\farcs 419$ ($79.8^\circ$) & 0.167 & 4.5 & 168 & 0.5\\
C$^{18}$O $J=2$--1 & 219.5603541  & $0\farcs 293 \times 0\farcs 420$ ($79.8^\circ$) & 0.167 & 3.0 & 88 & 0.5\\
SO $J_N=6_5$--$5_4$ & 219.9494420 & $0\farcs 291 \times 0\farcs 422$ ($80.2^\circ$) & 0.167 & 4.4 & 152 & 0.5 \\
H$_2$CO $J_N= 3_{0,3}$--$2_{0,2}$ & 218.2221920 & $0\farcs 291 \times 0\farcs 419$ ($82.2^\circ$) & 1.340 & 1.0 & 9.5 & 0.5
\enddata
\end{deluxetable*}

We requested observations for Oph~IRS63 as part of the eDisk Large program (project code: 2019.1.00261.L, PI N. Ohashi), however, our long-baseline data was not observed. Thus, we only present short-baseline molecular observations for Oph IRS63 obtained through ALMA DDT observations (2019.A.00034.S: PI: J. Tobin), and complement them with long-baseline continuum-only observations obtained from the ALMA archive (project code: 2015.1.01512.S, PI: D. Segura-Cox). The DDT observations were carried out on June 14, 2022, in configuration C-5 with baselines ranging from 15~m to 1.6~km with an on-source integration time of $\sim 30$~minutes that allowed us to obtain line data with angular resolution down to $\sim0.4\arcsec$. We observed $^{12}$CO (2-1) at a velocity resolution of $0.635$~km s$^{-1}$, while $^{13}$CO (2-1), C$^{18}$O (2-1), and SO (6$_5$-5$_4$)  were observed at a velocity resolution of 0.167~km s$^{-1}$. Two of the basebands were chosen to obtain the 1.3 mm continuum emission. 

The calibration and imaging tasks utilized the Common Astronomy Software Applications (CASA) package \citep{McMullin2007} version 6.2.1. A comprehensive description of the data reduction process, the spectral setup, and the calibrators can be found in \cite{Ohashi2023}. Table \ref{tab:observation_info} summarizes the observational parameters for continuum and line emission.

\section{Observational Results} \label{sec:results}

\begin{figure*}[ht!]
\centering
\epsscale{1.0}
\plottwo{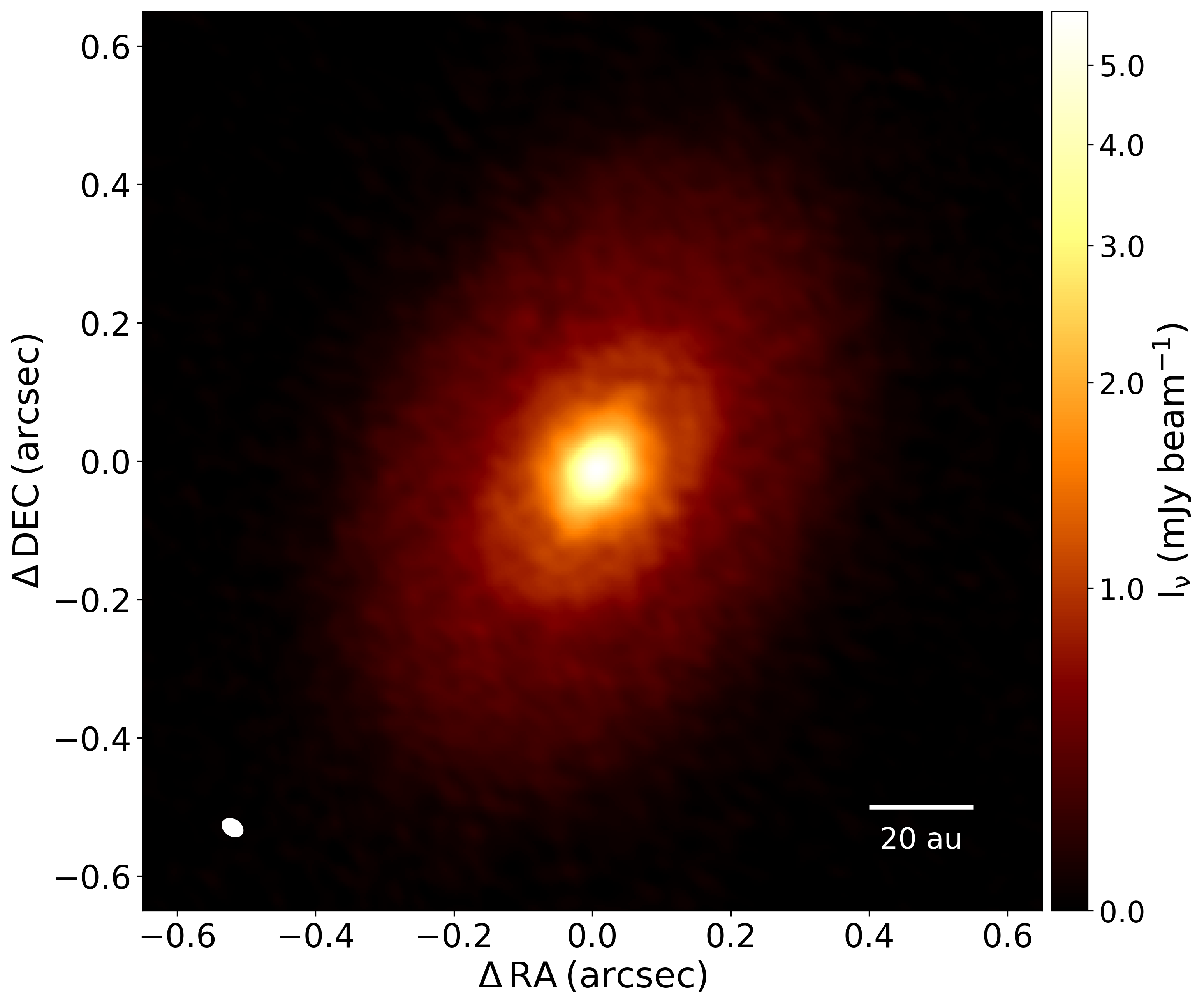}{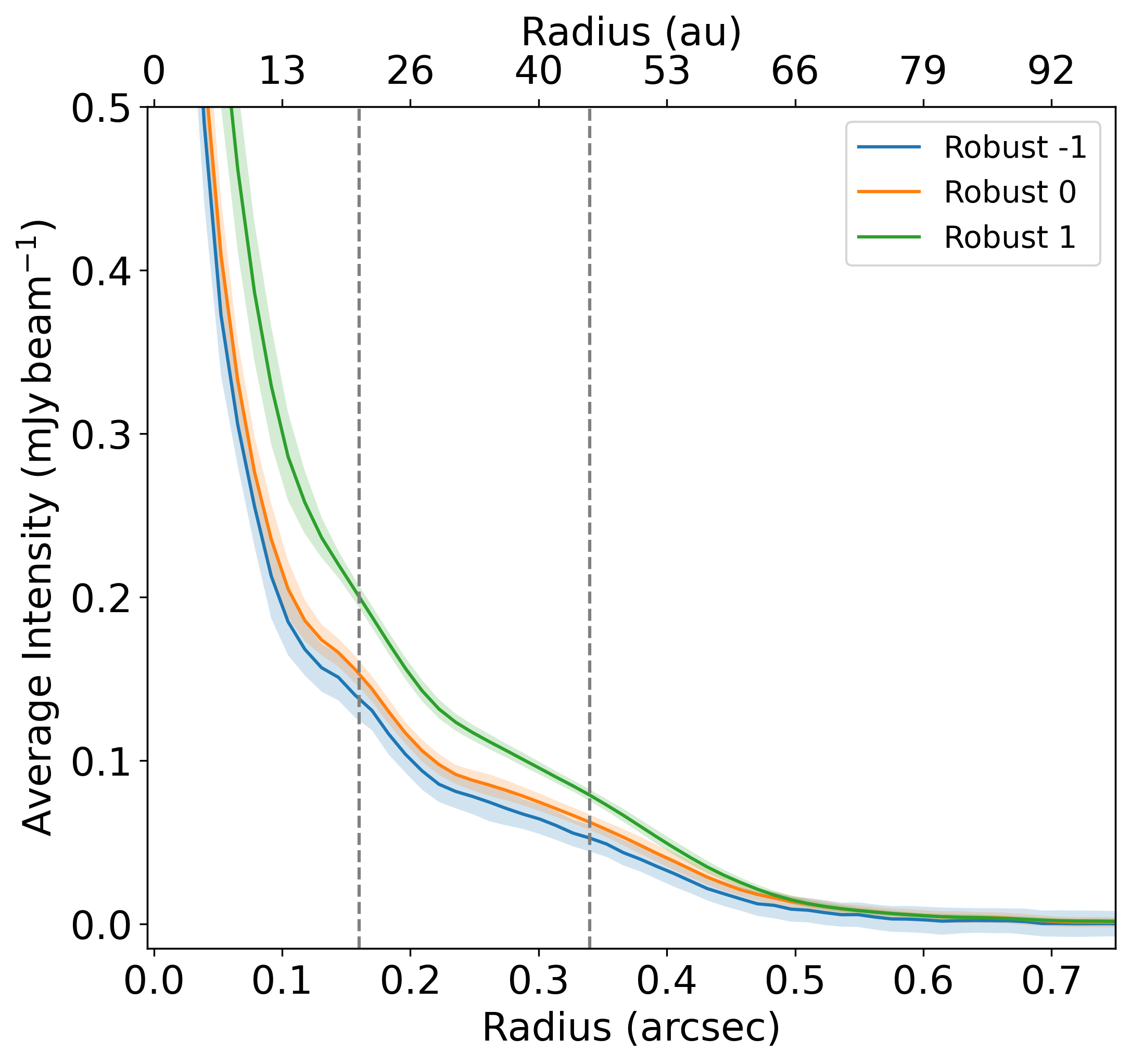}
\caption{Left: Dust continuum image with robust parameter R=0.0. The beam of size $0\farcs034 \times 0\farcs025$ is represented by a white ellipse on the bottom-left corner. A 20~au scale bar is shown for reference on the bottom-right corner. Right: Average radial intensity profile. Color lines represent radial profiles obtained for images with different Robust parameters. Two ring-like structures are visually identified at distances of $0\farcs16$ and $0\farcs34$, whose positions are in agreement with results from \cite{Segura-Cox2020}. 
}
\label{fig:continuum_image}
\end{figure*}

\subsection{Millimeter continuum} \label{subsec:continuum}

The 1.3~mm continuum image of the disk around Oph IRS63 is shown in Figure \ref{fig:continuum_image} (left panel). We see a disk that is centrally peaked with a peak intensity of 5.8 \mjbeam or 170 K. We fit a 2D Gaussian profile using the \textit{imfit} function in CASA to measure the basic dust disk properties \citep[see also][]{Ohashi2023}. We obtained a disk position angle $\rm P.A= 149^\circ$ (measured from north through east), a central position at coordinates RA =$16:31:35.654$, DEC = $-24:01:30.08$, and a deconvolved disk aspect ratio of $0.685$ that we transform to disk inclination of $i_{disk} = 46.7^\circ$. Because the Gaussian fit does not properly account for the extended dust continuum structure, we integrate the flux above 5$\sigma$ to derive a total flux density of $317\pm 0.5$ mJy.

To measure the dust disk size, we adopted the now common definition of the radius enclosing a certain fraction of the total flux \citep[e.g.,][]{Sanchis2021}. We used GoFish \citep{GoFish} to compute an azimuthally averaged flux (assuming inclination of $i=47^\circ$ and position angles of $\rm P.A.=149^\circ$), and defined the radii enclosing the 68$\%$, 90$\%$ and 95$\%$ of total flux as $R_{68\%}$= $0\farcs17$, $R_{90\%}$= $0\farcs38$, and $R_{95\%}$= $0\farcs47$, respectively. The difference in disk size at different flux cuts is due to the strongly peaked central emission observed for Oph~IRS63. 

From Figure \ref{fig:continuum_image}, we see that beyond the central peak, there is at least one diffuse annular structure. Such annular structures around Oph~IRS63, were studied in detail by \cite{Segura-Cox2020}, where two ring-like structures were identified at $R\sim 0\farcs19$ and $R\sim 0\farcs35$ with widths of $0\farcs04$ and $0\farcs09$, respectively. On the right panel of Figure \ref{fig:continuum_image}, we show azimuthally averaged intensity profiles of the continuum emission for different robust parameters. We visually confirm the presence of an inner ring at $\sim 0\farcs16$, and an outer ring at $0\farcs34$. In this work, we do not intend to perform any detailed analysis of the continuum emission, as a future paper will address the substructures and asymmetries observed in all the sources of the eDisk sample.

We calculate the mass of the dust disk following equation (\ref{eq:eq_dust_mass}) proposed by \cite{Hildebrand1983}, which assumes that the emission is optically thin and comes only from a thermal component. In this equation, $S_{\nu}$ is the flux density calculated by integrating the emission above $5\sigma$, D is the distance to the source of 132~pc, $B_{\nu}(T_{d})$ is the Planck function at the observed frequency (225 GHz), $T_{d}$ is the dust temperature, and $\kappa_{\nu}$ is the absorption coefficient adopted as $\kappa_{\nu}$ = 2.3 cm$^{2}$ per gram of dust \citep{Beckwith1990}. 

\begin{equation}
\label{eq:eq_dust_mass}
      M =  \frac{S_{\nu} D^{2}}{ \kappa_{\nu}B_{\nu}(T_{d})},  
\end{equation}

We use two different dust temperatures to measure the mass dust of the disk: 1) the standard $ \rm T_{dust} = $ 20~K value often used in Class II sources \citep[e.g.,][]{Pascucci16}, from which we obtain a total dust disk mass of $\rm M_{dust}\sim 5 \times 10^{-4} \, M_\odot$ (or $\rm M_{dust} \sim 160 \, M_\oplus$). 2) the temperature prescription proposed by \cite{Tobin2020} of $\mathrm{T_{dust}= 43 \, (L_{bol})^{1/4}}$, which for Oph~IRS63 yields a temperature $\rm T_{dust} = $46~K, and therefore a dust mass of $\rm M_{dust}\sim 2 \times  10^{-4} \, M_\odot$ (or $\rm M_{dust} \sim 60 \, M_\oplus$). These values should be regarded as lower limits for the total disk mass of solids since the 1.3 mm emission in the disk is optically thick as discussed in \cite{Segura-Cox2020}.



\subsection{Molecular tracers} \label{subsec:lines}

\begin{figure*}[ht!]
\centering
\epsscale{1.15}
\plotone{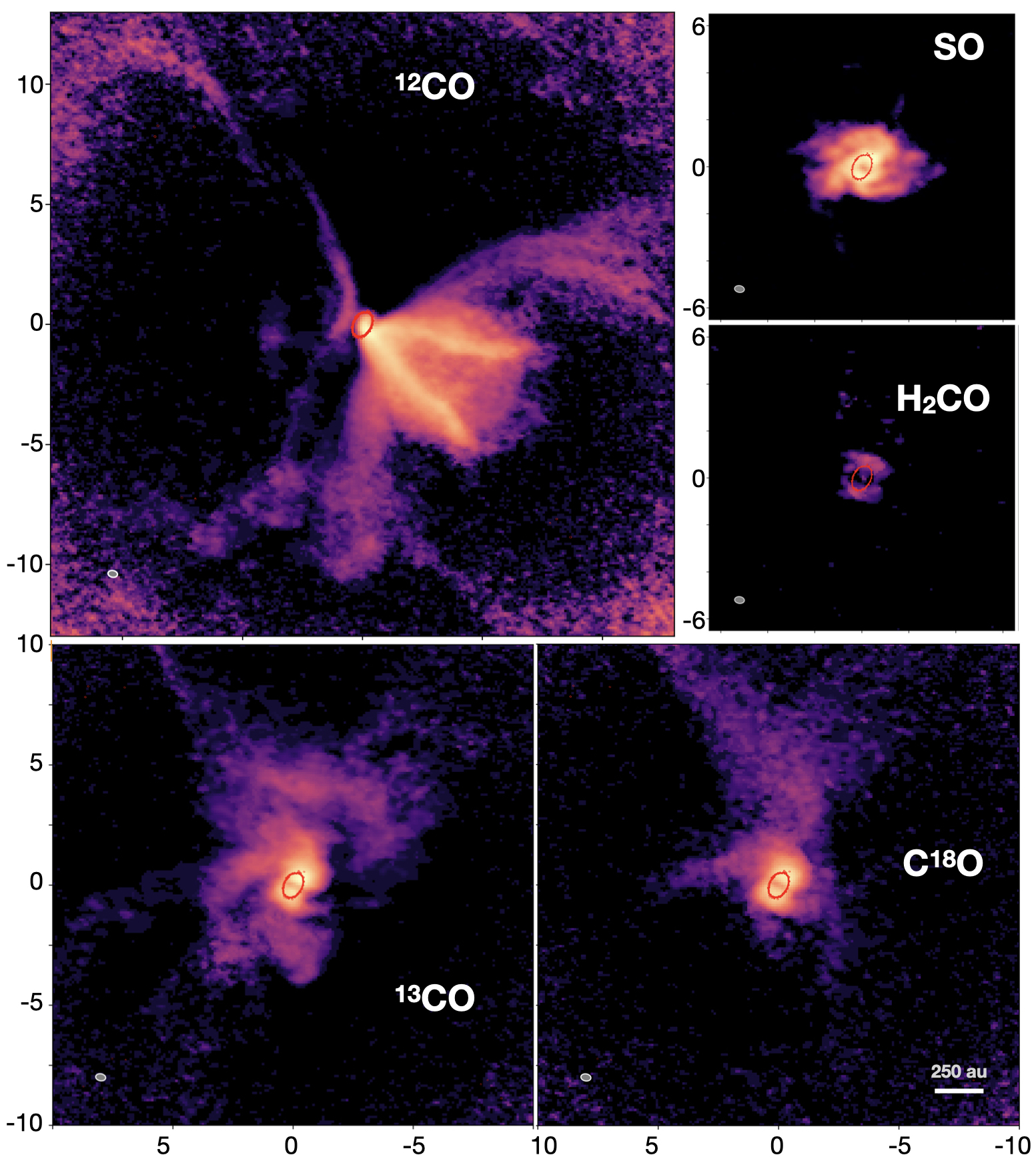}
\caption{Integrated intensity maps of the molecules presented in this work. The emission goes from 1$\sigma$ of the integrated intensity to the maximum value of the map. The color scales were stretched using an $arcsinh$ function to emphasize faint structures. The 5$\sigma$ contour of the continuum emission is ploted in red. For the $^{12}$CO molecule, the map size is $26\arcsec \times 26\arcsec$ and the velocity integration are -12.37 to -0.94 \kms and 5.40 to 17.48 \kms. For $^{13}$CO and C$^{18}$O, the map sizes are $20\arcsec \times 20\arcsec$. The moment 0 maps for $^{13}$CO and C$^{18}$O were integrated between -2.33 to 1.85 \kms and 3.18 to 6.02 \kms; and 0.01 to 2.35 \kms and 2.85 to 5.69 \kms, respectively. For SO and H$_2$CO, the maps sizes are $13\arcsec \times 13\arcsec$. The velocity integrations are 0.18 to 5.52 \kms, and 0.72 to 4.72 \kms, respectively. Beam sizes are depicted at the bottom left of the maps, according to Table \ref{tab:observation_info}. A horizontal bar of size 250~au is presented on the bottom right of the C$^{18}$O map. All maps are centered using the Gaussian fit to the continuum emission at the coordinates R.A. = 16:31:35.654 and decl.=-24:01:30.08.  
}
\label{fig:mosaic_of_molecules}
\end{figure*}

In Figure \ref{fig:mosaic_of_molecules}, we present a gallery of the detected molecular tracers described in this work. The images correspond to integrated intensity maps (moment 0) of three CO isotopologues: $^{12}$CO, $^{13}$CO, C$^{18}$O, in addition to the SO and H$_2$CO lines. The different tracers reveal a large-scale envelope, a compact disk, spiral-like features, and an outflow. Although some lines trace multiple structures, below, we attempt to present the different components of the protostellar system, focusing on the most important tracers for each case.

In some molecular tracers, our observations suffer from cloud contamination at channel maps close to systemic velocity. This feature is most likely caused by gas being filtered out by the ALMA interferometer. Additionally, we see a strong negative absorption at the disk position for all the molecular species (except in SO). This artifact is caused by continuum over-subtraction, resulting from optically thick molecular lines at the line center, and it must be kept in mind when discussing some molecular maps close to the stellar position (see channel maps in Appendix \ref{appendix:channel_maps}). 

\subsection{Disk}
\label{subsec:results_disk_emission}
\begin{figure*}[ht!]
\centering
\epsscale{1.15}
\plotone{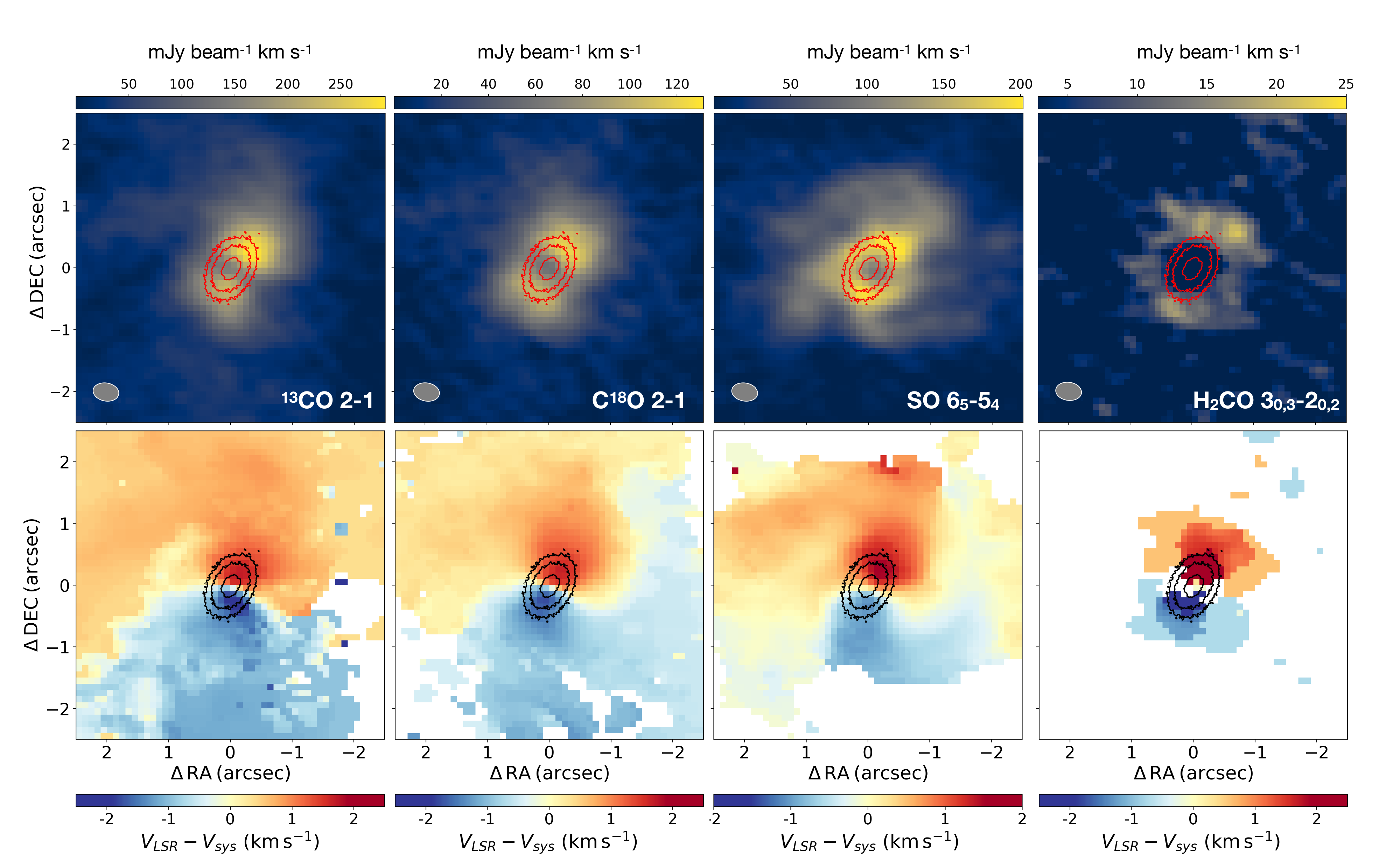}
\caption{Comparison between the continuum (contours) and molecular emission of $^{13}$CO, C$^{18}$O, SO, and H$_2$CO (colors), respectively. The continuum emission is plotted at contours levels of 6$\sigma$, 20$\sigma$, and 50$\sigma$. The beam size of the continuum is 0$\farcs034 \, \times \, 0\farcs025$ while the beam size of the gas is $\sim$10 times larger. The top row shows the Moment 0 maps, while the bottom row shows the Moment 1 maps. The velocities of the moment 1 maps are corrected for $\vsys$ (i.e., we subtracted $\vsys$ = $2.8$~$\rm km \, s^{-1}$ to the map). The FOV of all the maps is $5\arcsec \times 5\arcsec$.}
\label{fig:Disk-size-comparison}
\end{figure*}

Following the results of the dust continuum, we now focus on the gas emission at or close to the dusty disk position. Figures \ref{fig:mosaic_of_molecules} and \ref{fig:Disk-size-comparison} show three prominent tracers of gas at the disk position: $^{13}$CO, C$^{18}$O and SO. The H$_2$CO line only shows weak emission, which is strongly affected by continuum over-subtraction.

Figure \ref{fig:Disk-size-comparison} shows zoom-in maps of the molecular tracers centered around the disk position. The moment 0 maps of $^{13}$CO and C$^{18}$O show emission mostly elongated along the dust disk major axis and their moment 1 maps show a clear velocity gradient along this direction. The SO and H$_2$CO moment 0 maps (shown in Figure \ref{fig:Disk-size-comparison}) show a velocity gradient along the direction of the disk major axis, but they do not show an elongation on the moment 0 maps. Instead, they show a complex pattern with an apparent lack of emission at the center of the map. Although the lack of central emission in H$_2$CO can be attributed to continuum over-subtraction (see Figure \ref{fig:channel_maps_H2CO} of Appendix \ref{appendix:channel_maps}), the ring-like pattern in SO may be real and is further described in Section  \ref{subsec:results_Spirals}.




The velocity gradients along the major axes and the elongated emission (in the same direction as the dust disk major axis) in $^{13}$CO and C$^{18}$O suggest the presence of a rotating disk-like structure around Oph~IRS63. The size of the gas disk and the exact velocity pattern of the rotation, e.g., if it is consistent with Keplerian rotation or not, are further analyzed in Section \ref{subsec:SLAM_analysis}. 

In these close-in maps, we also observe an S-shaped pattern at the transitional zone between blue and red-shifted emission in C$^{18}$O and SO. This S-shape depicts a change in the direction of the velocity gradient between an outer region ($\gtrsim1\arcsec$) and the inner region ($\lesssim 1\arcsec$) which aligns with the dust disk's position angle. This change in velocity gradient is a clear example of an infalling envelope connecting with a rotating disk structure \citep[as shown previously in e.g.,][]{Momose1998,Murillo2013,Yen2013}.

\subsection{Envelope}
\label{subsec:results_envelope_emission}

\begin{figure*}[ht!]
\centering
\epsscale{1.1}
\plottwo{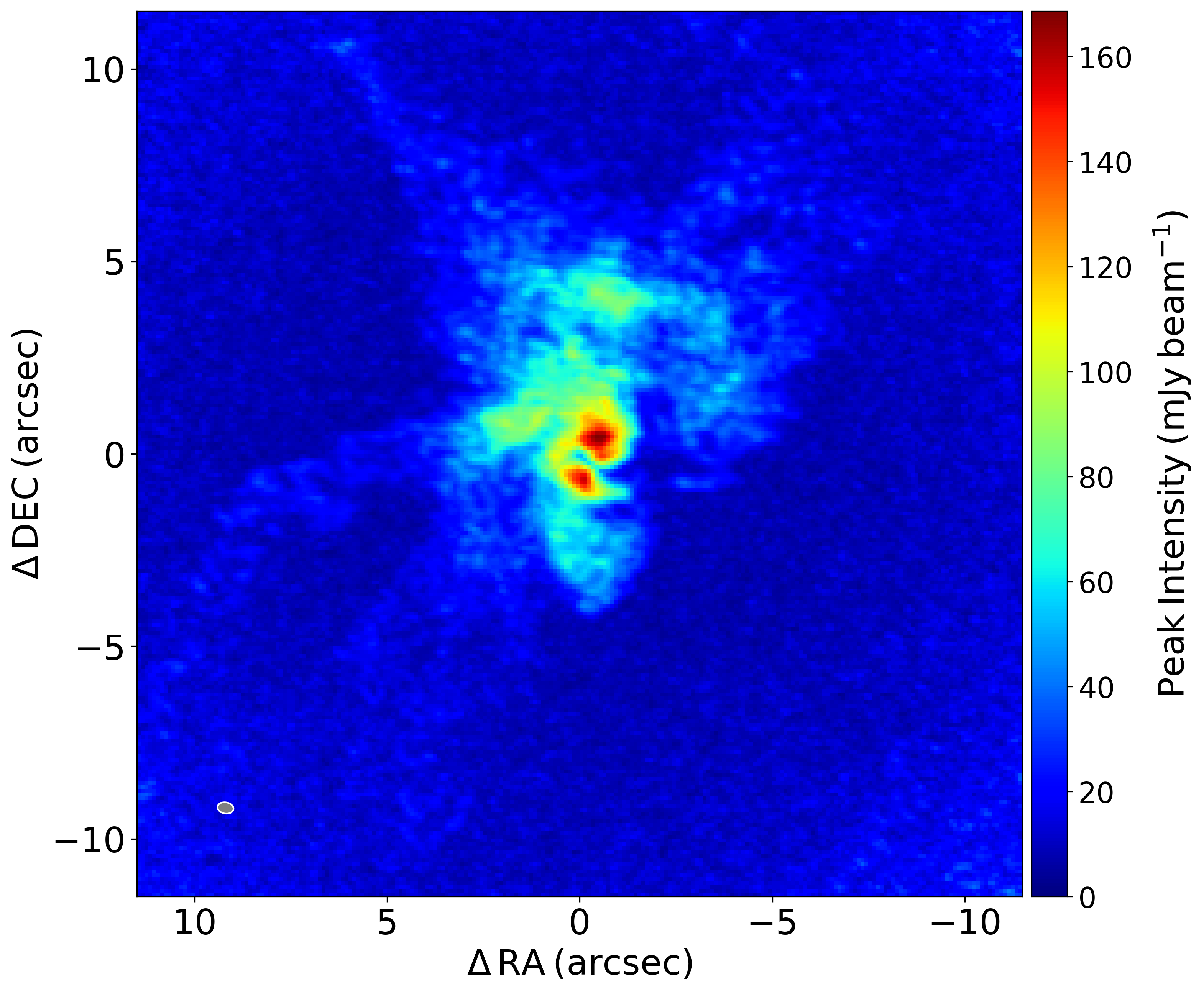}{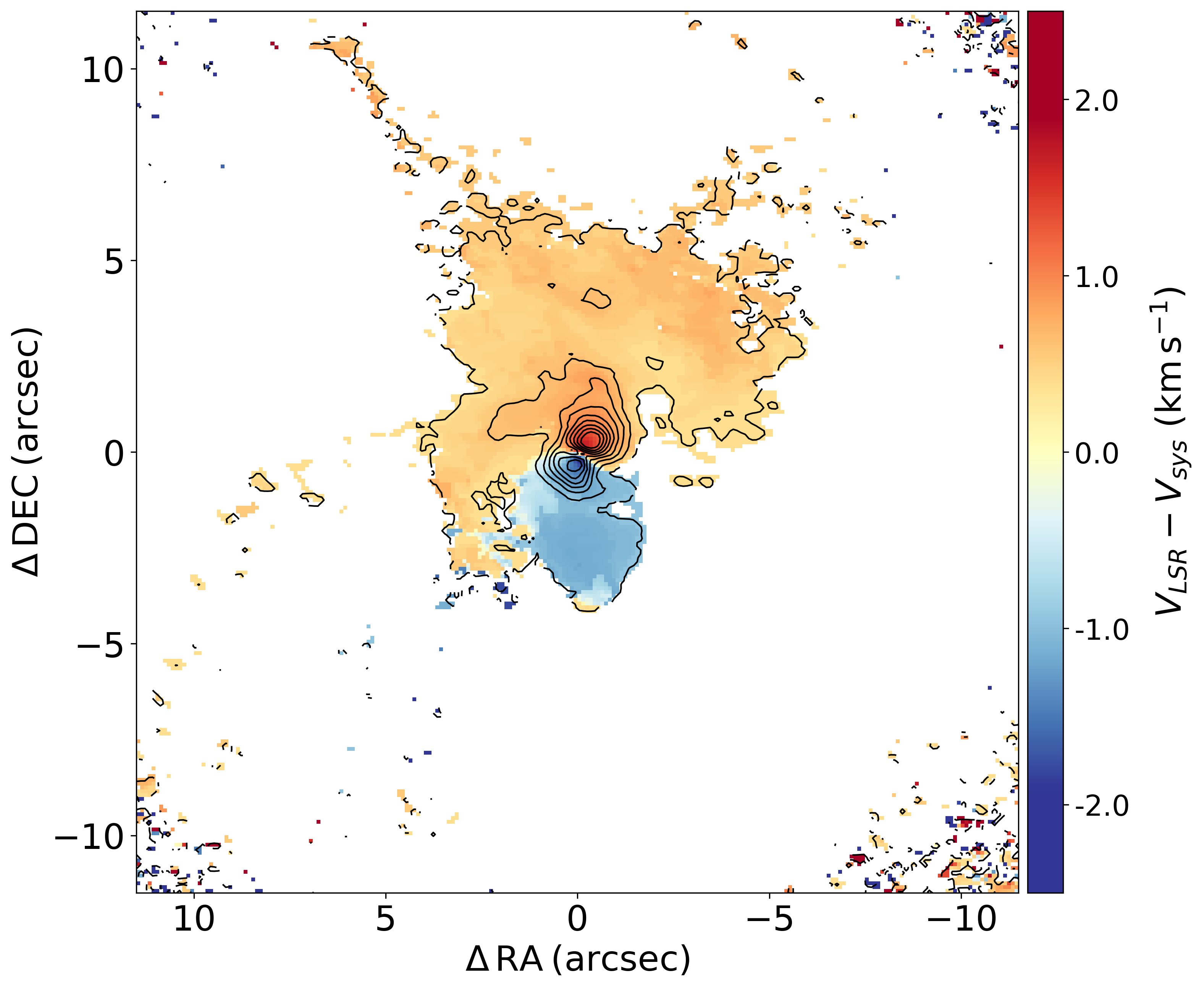}
\caption{Large-scale maps of $^{13}$CO. The left plot shows the peak intensity map (moment 8) in units of \mjbeam. The right plot shows the velocity structure (moment 1 map) centered at $V_{sys}=$2.8~$\rm km \, s^{-1}$. The moment 1 map was obtained by integrating emission above 22.5 mJy beam$^{-1}$ (5 times the rms). The integrated intensity (moment 0) is overlaid in black contours starting at $1\sigma$ ($1\sigma$=4.7 \mjbeamkms) with a step of $6\sigma$. Both maps show the connection of a large-scale structure ($>5 \arcsec$) to a compact ($\sim 2 \arcsec$) disk-like structure. The field of view in both maps is $23\arcsec \times 23\arcsec$. 
}
\label{fig:13CO_map}
\end{figure*}

From the gallery plot of Figure \ref{fig:mosaic_of_molecules}, we see that a large envelope structure is detected in $^{13}$CO and C$^{18}$O, which extends to distances greater than 10 arcseconds ($\sim1300$~au). In fact, from Figure \ref{fig:13CO_map}, we see $^{13}$CO extended asymmetric emission reaching $\sim13\arcsec$ from the stellar position. While most of the extended emission is relatively faint, it is still detected at levels of $\sim22$~\mjbeam (S/N$\sim$5). The large-scale structure of the $^{13}$CO is highly complex, with several arcs/arms features extending to the northwest, northeast, and southeast. In contrast, towards the southwest direction, there is a lack of extended structure, which coincides with the position where the $^{12}$CO blue-shifted outflow is observed. The large-scale velocity structure of $^{13}$CO (right panel in Figure \ref{fig:13CO_map}) shows a velocity gradient mainly in the north-south direction, with most of the red-shifted emission in the north and the blue-shifted emission in the south.

\begin{figure*}[ht!]
\centering
\epsscale{1.1}
\plottwo{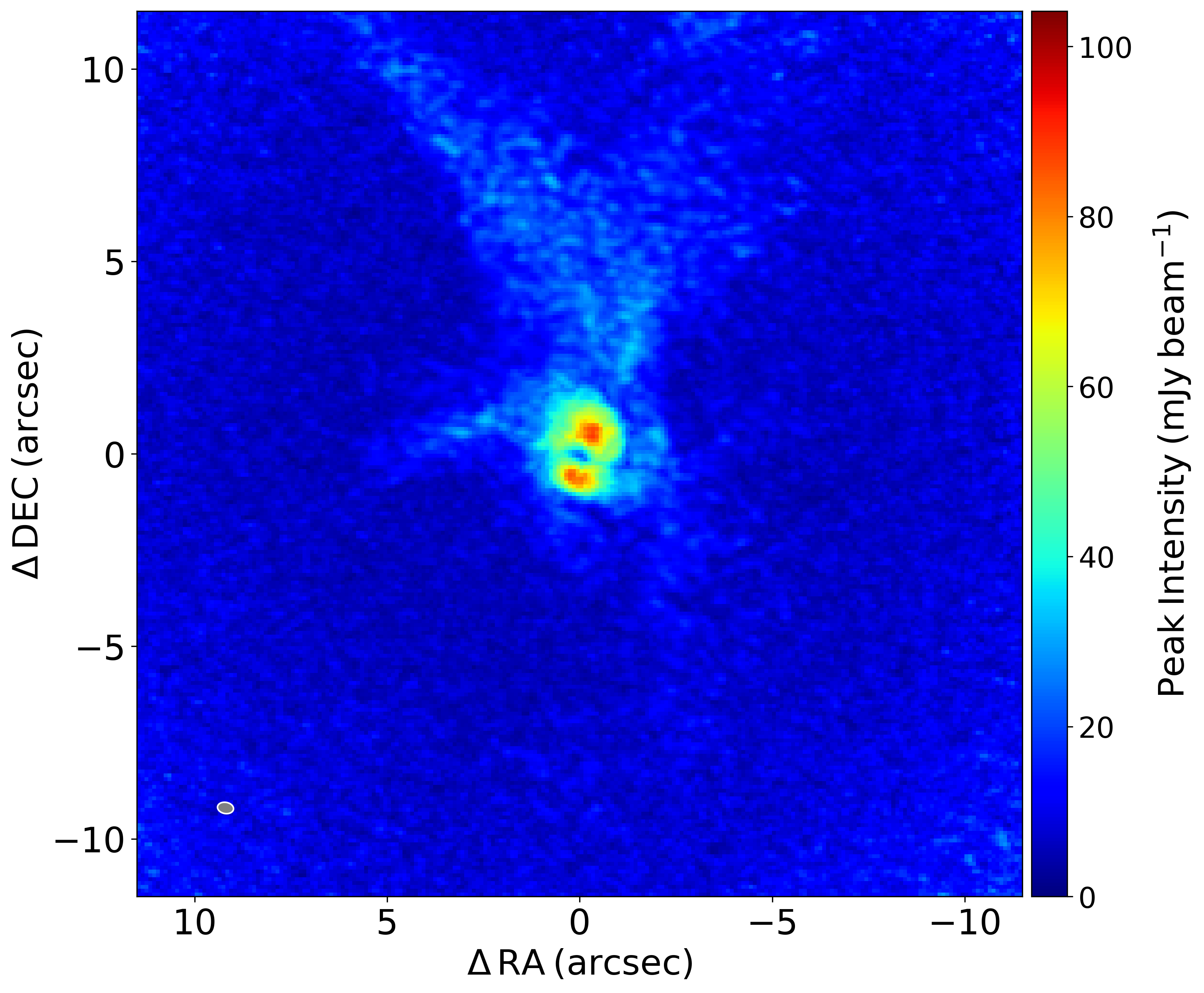}{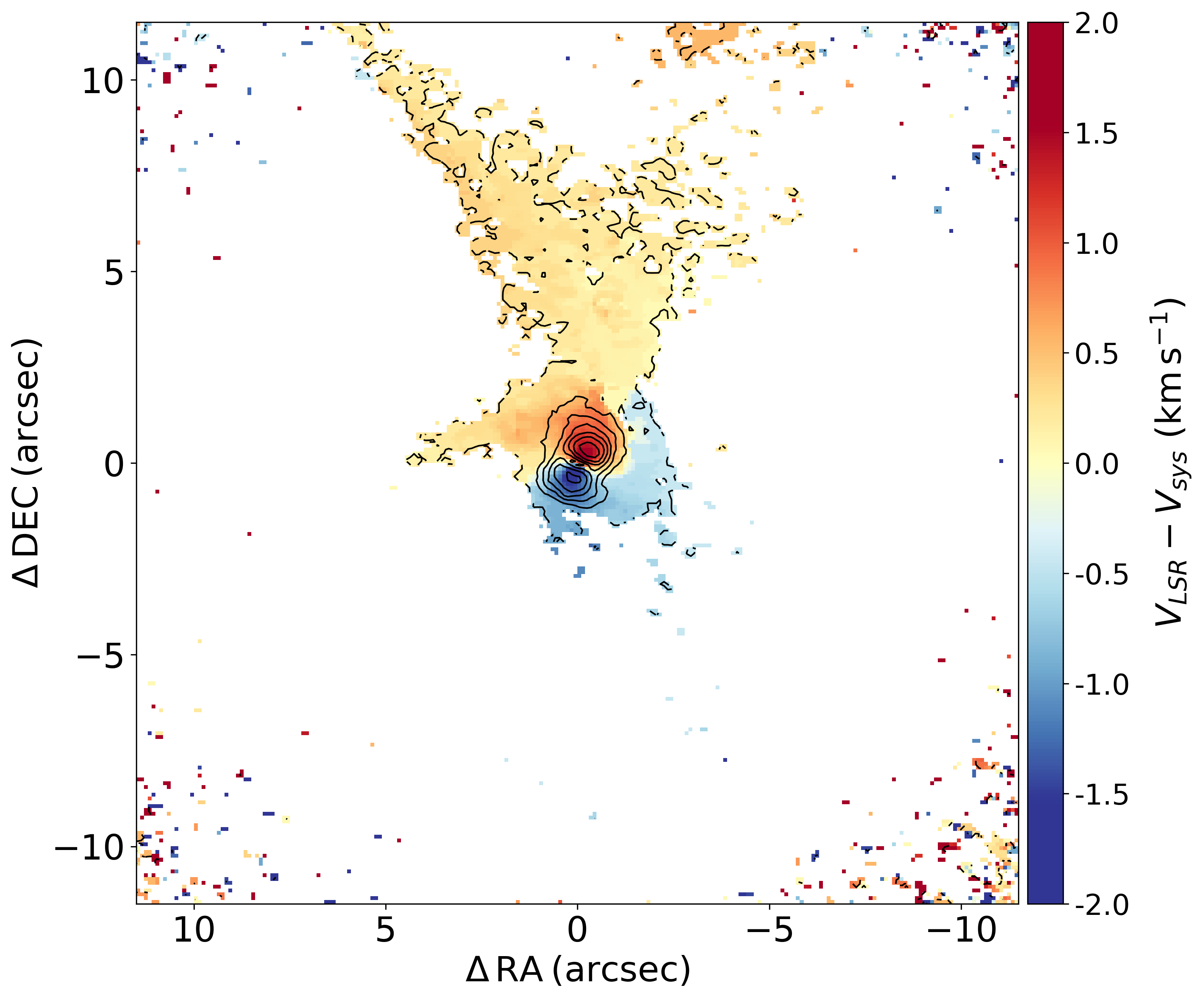}
\caption{Large-scale maps of C$^{18}$O. The left plot shows the peak intensity map (moment 8) in units of $\rm mJy \, beam^{-1}$. The right plot shows the velocity structure (moment 1 map) centered at $V_{sys}=$2.8~$\rm km \, s^{-1}$. The moment 1 map was obtained by integrating emission above 15 $\rm mJy \, beam^{-1}$(5 times the rms). The integrated intensity (moment 0) is overlaid in black contours starting at $1\sigma$ ($1\sigma$=2.3 \mjbeamkms) with a step of $6\sigma$. The field of view in both maps is $23\arcsec \times 23\arcsec$.}
\label{fig:C18O_map}
\end{figure*}

Figure \ref{fig:C18O_map} shows the moment 8 (maximum value in the spectrum) and the moment 1 maps for the C$^{18}$O line. Similar to the $^{13}$CO line, we see extended emission reaching $\sim13\arcsec$, but in the case of  C$^{18}$O, the extended emission is mostly detected towards the northern part of the system. In fact, a large arc-like feature is evident from the moment 8 map, which branches off to the northeast and northwest. In addition, a small arc structure ($\sim5\arcsec$ in length) is detected in the eastern part at levels of $\sim$20-40 \mjbeam.


In both CO isotopologues, the northern emission (red-shifted) is brighter and more extended, consistent with our interpretation that the northern is the near side of the system, and the southern part is partially hidden behind the large-scale $^{12}$CO outflow (see Section \ref{subsec:schematic_view_system}). The observed north-south brightness asymmetry could also be caused by the large number of channels missed due to cloud contamination, which affect more strongly the $^{13}$CO as more emission is being resolved out.

Finally, when considering the orientation of the $^{13}$CO and C$^{18}$O envelopes, we find that they have velocity gradient directions roughly aligned North-South, as measured from the peak emission in moments 1 and 8 maps. The orientation of these large-scale velocity gradients, however, differs from what is observed at smaller scales ($\lesssim1\arcsec$), where the velocity gradient aligns with the position angle of the dust disk (i.e., $\rm P.A= 149^\circ$, as shown in Figure \ref{fig:Disk-size-comparison}). Similar to the S-shape zero-velocity line in SO, the change in velocity gradient direction between large- and small-scale in $^{13}$CO and C$^{18}$O is indicative of an infalling envelope \citep[e.g.,][]{Sakai2016}.

\subsection{Spirals}
\label{subsec:results_Spirals}



\begin{figure}[ht!]
\centering
\epsscale{1.15}
\plotone{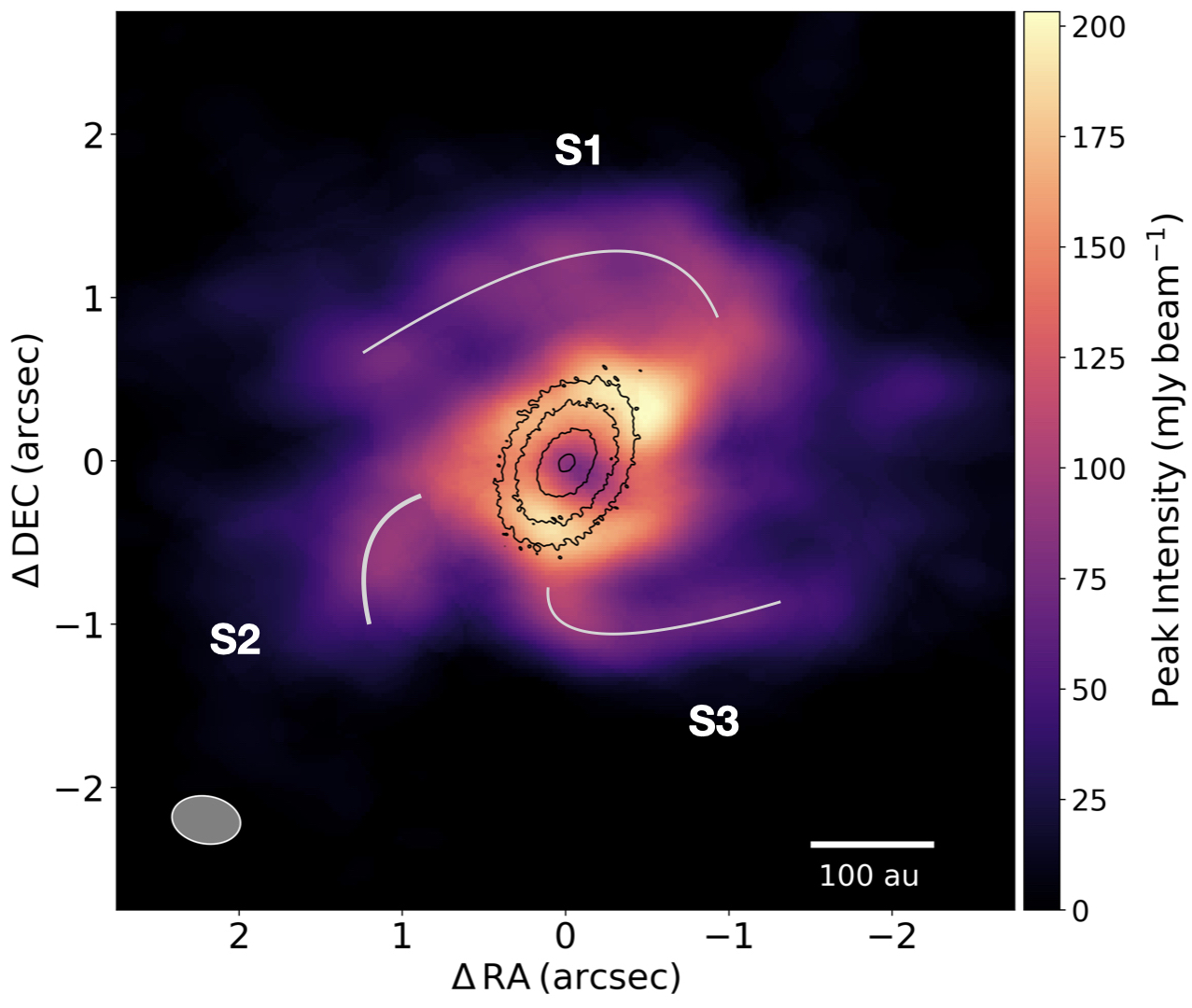}
\caption{Integrated intensity of the SO emission to highlight the spiral pattern. The moment 0 map was generated using only emission above 3$\sigma$. The spirals S1, S2, and S3 depict the spiral patterns connecting with the ring-like feature on the north, east, and south, respectively. The continuum is shown in black contours at 6$\sigma$, 20$\sigma$, 40$\sigma$, and 150$\sigma$. The image has a field of view of $5 \farcs5 \times 5\farcs5 $}.
\label{fig:SO_spirals_annotation}
\end{figure}

One of the most striking observational result for Oph~IRS63 is the detection of a spiral-like structure in the SO $6_5 - 5_4$ tracer. Figure \ref{fig:SO_spirals_annotation} shows the SO integrated intensity map, with annotations marking the spiral-like structures that connect at the north (S1), east (S2), and south (S3) with the disk. The apparent motion of the three spirals is as if they were rotating clockwise. The S1 and S3 spirals are tightly wound around the disk, while S2 extends outwards towards the south. 
The three spirals connect to a ring-like structure in the inner $0\farcs5$ of the SO emission. The S1 spiral connects with the annular structure at a P.A. of $\sim320^\circ$ where the gas has its maximum intensity at about 200~\mjbeamkms. The S2 spiral connects with the ring at a P.A. of $\sim90^\circ$ (almost exactly east), but it does not produce an emission peak. The S3 spiral connects with the ring at a P.A. of $\sim170^\circ$, where the gas emission reaches about 190~\mjbeamkms.  

Figure \ref{fig:Disk-size-comparison} shows the SO velocity map, which exhibits a dominant velocity gradient along the major axis of the dust disk, similar to that found for $^{13}$CO and C$^{18}$O at distances $\lesssim 1\arcsec$, but it becomes significantly distorted with increasing radius (leading to the S-shape zero velocity line). First, the S3 spiral coincides with the blue-shifted velocity pattern, while the S1 spiral traces an opposite structure with a red-shifted velocity. The S2 spiral starts close to the disk at a positive but almost zero velocity offset and then becomes blue-shifted at increasing radial distances. From the channel maps presented in the Appendix (Fig \ref{fig:channel_maps_SO}), we can see that most of the spiral structures appear close to the systemic velocity at values of $\pm$1.3~$\rm km \, s^{-1}$ with respect to $V_{sys}$=2.8~$\rm km \, s^{-1}$. These velocities ranges are similar to the extended envelope structure identified in $^{13}$CO and C$^{18}$O, and significantly different from the outflow detected in $^{12}$CO.

As briefly mentioned above, an annular structure with a radius of $\sim0.5\arcsec \pm 0.2\arcsec$ is observed on the moment 0 map, and this ring-like feature peaks close to the edge of the dust disk (see Section \ref{subsec:continuum}). The center of the SO ring, or more precisely, the center of the emission drop, is offset by approximately $0\farcs11$ relative to the dust continuum emission, as measured through a 2D Gaussian fitting (see Figure \ref{fig:SO_spirals_annotation}). Although small, this difference is significant given that we can measure the positions of the SO emission with an accuracy of $0\farcs023$\footnote{Based on the ALMA Technical Handbook estimates for absolute nominal positional accuracy. Observing conditions could affect the actual positional accuracy by a factor of two compared to the nominal value.} 


In the literature, different SO transitions can trace various components of a protostar, including jets, low-velocity outflow, warm inner envelope, and disk emission \citep[e.g.,][]{Tychoniec2021}. Due to the kinematics and spatial distribution of SO shown, we can rule out an outflow origin for the case of Oph IRS63. Instead, given the larger rotating envelope structure detected in $^{13}$CO, and the prominent S-shape pattern at the transitional zone between blue- and red-shifted emission, we speculate that it comes from the connection between the warm inner envelope and the disk. This idea is further discussed in Section \ref{sec:disc_SO_spirals}.

A different type of arc or spiral-like structures are found in the C$^{18}$O moment maps (shown in Figure \ref{fig:C18O_map}). These are large-scale features ($>5\arcsec$), have lower pitch angles than the SO (less tightly wound), and can definitely be associated with the envelope structure. Three to four of these structures can be identified from the moment maps of C$^{18}$O (see Figure \ref{fig:C18O_map}), but the most conspicuous ones are the northern and the eastern ones extending $\sim13\arcsec$ and $\sim5\arcsec$, respectively. Section \ref{sec:Analysis} explores the idea that these structures result from an infalling rotating envelope that feeds the disk with material from the collapsing cloud.  

\subsection{The Outflow}
\label{subsec:results_outflow}

\begin{figure*}[ht!]
\centering
\epsscale{1.1}
\plotone{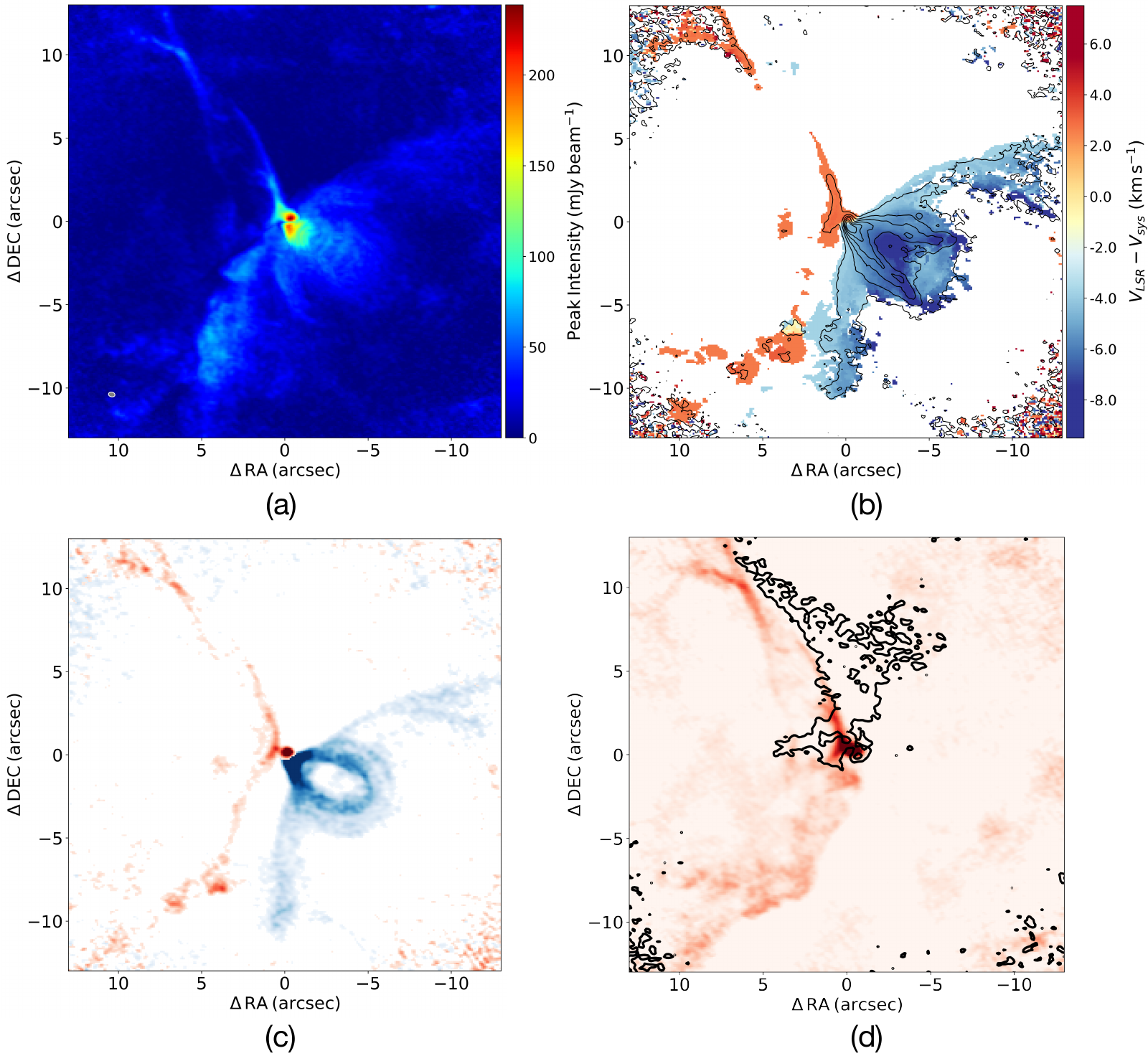}
\caption{Four maps showing different aspects of the $^{12}$CO outflow emission. Panel (a) shows the peak intensity or moment 8 map. Panel (b) shows the moment 1 map in colors, obtained by integrated emission above 6$\sigma$, from channels at $V_{LSR}$ =-12.37 to -0.94~$\rm km \, s^{-1}$ and from  $V_{LSR}$ =4.78 to 17.48~$\rm km \, s^{-1}$. The map is also overlaid in black contours by the moment~0 map starting at $3\sigma$, in steps of $8\sigma$ ($1\sigma = 8$~\mjbeamkms). 
In panel (c), we show two channels at $V_{LSR}$ =-2.21~$\rm km \, s^{-1}$ (blue-shifted side) and $V_{LSR}$=6.05~$\rm km \, s^{-1}$ (red-shifted side) to emphasize the complex shell-like structure of the outflow. Panel (d) is an overlay between red-shifted outflow at $V_{LSR}$=4.78~$\rm km \, s^{-1}$ and the arc-like emission of C$^{18}$O from channel at $V_{LSR}$ = 3.02 $\rm km \, s^{-1}$ (in black contours). The FOV in all maps is $26\arcsec \times 26\arcsec$}
\label{fig:outflow}
\end{figure*}

In Figure \ref{fig:outflow}, we plot the $^{12}$CO emission around Oph~IRS63, revealing a large-scale bipolar outflow structure. 
The outflow extends to the northeast and southwest direction until the edge of the map, where the sensitivity steadily decreases due to the poorer primary beam response. Panels (a) and (b) of Figure \ref{fig:outflow} display a strong and blue-shifted outflow emission towards the southwest. Conversely, the northeast outflow is red-shifted and weaker, appearing only in a few channel maps. A tentative detection of the inner $\sim 1\arcsec$ of this outflow was recently reported in \cite{Antilen2023}.

The blue-shifted side of the outflow exhibits shell-like structures, which can be roughly categorized into narrow and broad components. The narrow component is clearly detected up to velocities of $-15$~$\rm km \, s^{-1}$, and it forms a closed loop-like structure with a maximum radial size of $\sim8\arcsec$ at velocities of $-3.0$ to $-7.0\,\rm km \, s^{-1}$. In the moment zero map (see Figure \ref{fig:mosaic_of_molecules}), the narrow blue-shifted component displays a characteristic V-shape pattern. We estimated the outflow's position angle to be $\sim 240^\circ$ by averaging the P.A.s of the two legs of the V-shape structure.

The broad component is detected at approximate velocities of $ -3$ to $ -9.0~ \rm \,  km \, s^{-1}$. The latter always fully encompasses the narrower component, and its opening angle decreases with increasing negative velocity. A similar configuration, in which a broader outflow fully contains a narrower component, has been observed in ALMA observations of DO Tau \citep{Fernandez-Lopez2020}, DK Cha \citep{Harada2023}, and GSS30 IRS3 \citep{Santamaria-Miranda2023}. Near-infrared observations of such configuration have also been reported, for example, for HL Tau \citep{Takami2007}. Furthermore, the narrower blue-shifted component in Oph~IRS63 (and marginally also the broader component) is split into even finer structures as shown in panel (c) of Figure \ref{fig:outflow}. 
A future study will determine whether the complex shell-like structure is a kinematic feature produced by different episodic events \citep[e.g.,][]{Zhang2019,Fernandez-Lopez2020} or whether a single outflow driven by magnetized winds and their interplay with the environment can produce the observed shells \citep{Shang2020,Shang2023}.

The red-shifted outflow is less prominent and more asymmetric, and can be more easily identified in panel (c) of Figure \ref{fig:outflow}, where selected channels are used to emphasize the large-scale $^{12}$CO emission. Unlike the blue-shifted side, the red-shifted outflow only shows a broad (large opening angle) component that appears to extend broadly towards the east direction. Although purely speculative, interaction with an inhomogeneous ambient medium may have modified the appearance of the red-shifted outflow direction. Alternatively, an outflow precession mechanism may be at play \citep[e.g.,][]{Vazzano2021}.



In panel (d) of Figure \ref{fig:outflow}, we overlaid the red-shifted component of the outflow with the northern C$^{18}$O arc-like emission, as these two appear spatially and morphologically close to each other. The northern C$^{18}$O emission appears to be at the outer edge of the $^{12}$CO outflow and follows its shape closely. Although we only access a two-dimensional projection of the complex system, the C$^{18}$O gas likely lies outside the cavity opened by the outflow. If this is the case, then when the outflow opened the cavity observed in $^{12}$CO, it must have pushed the ambient gas around creating a localized higher density and temperature region that we observe as this C$^{18}$O arc-like emission. \cite{Thieme2022} recently reported a similar streamer and outflow configuration for the Lupus 3-MMS source.

Lastly, the outflow morphology is related to the orientation of the disk. We can infer that the northeastern edge is the near edge of the disk based on the red-shift of the northeast outflow component relative to the source's systemic velocity. Additionally, using the outflow position angle of P.A. $\sim240^\circ$, we confirm that the blue-shifted outflow is perpendicular to the disk direction. However, the red-shifted outflow axis appears to be tilted with respect to the disk rotational axis.





\section{Analysis} 
\label{sec:Analysis}


\subsection{Kinematic analysis of the disk and envelope}
\label{subsec:SLAM_analysis}
Protostellar systems are typically identified as young stars surrounded by a disk and an envelope. The presence of an envelope around a young star distinguishes it from a more evolved Class II or T Tauri star. To determine if the compact ($<2\arcsec$) disk-like structure is a rotating Keplerian disk and whether the large ($\sim10\arcsec$) structure is an infalling envelope, we perform a kinematic analysis using the Spectral Line Analysis/Modeling (SLAM) code \citep{SLAM}.

SLAM fits  power law functions to rotational velocity profiles. This method takes a Position-Velocity (PV) diagram as input and fits a power law function (described by equation \ref{eq:slam}) to the data points. In this equation, $V_{m}(x)$ represents the velocity of the model as a function of the impact parameter $x$ (i.e., the distance coordinate in the PV diagram), $p_{in}$ is the  power law index of the inner part of the emission that changes to $p_{in}+dp$ at a breaking radius $R_b$ and breaking velocity $V_b$. $V_{sys}$ is the systemic velocity of the system. SLAM selects data points from the input PV in two ways; the ridge mode traces the peak of the emission at different channels, while the edge mode traces the outermost emission contour. SLAM performs the fitting process in an  MCMC fashion and it returns the best-fit parameters along with their formal uncertainties.

\begin{equation}
\label{eq:slam}
  V_{m} (x) =
    \begin{cases}
       sgn(x) V_b \left( \frac{|x|}{R_b} \right)^{-p_{in}} + V_{sys} & |x| \leq R_b \\
       sgn(x) V_b \left( \frac{|x|}{R_b} \right)^{-(p_{in}+dp)} + V_{sys} & |x| > R_b
    \end{cases}       
\end{equation}


It is worth mentioning that a Keplerian rotating structure has a rotational profile that falls off as the square root of the radius (i.e., $V \propto r^{-0.5}$), while an infalling envelope has a steeper drop in the rotational velocity with a profile following $V \propto r^{-1}$  \citep[e.g.,][]{Momose1998, Yen2013}.

\subsubsection{CO emission}

\begin{figure*}[ht!]
\centering
\epsscale{1.15}
\plotone{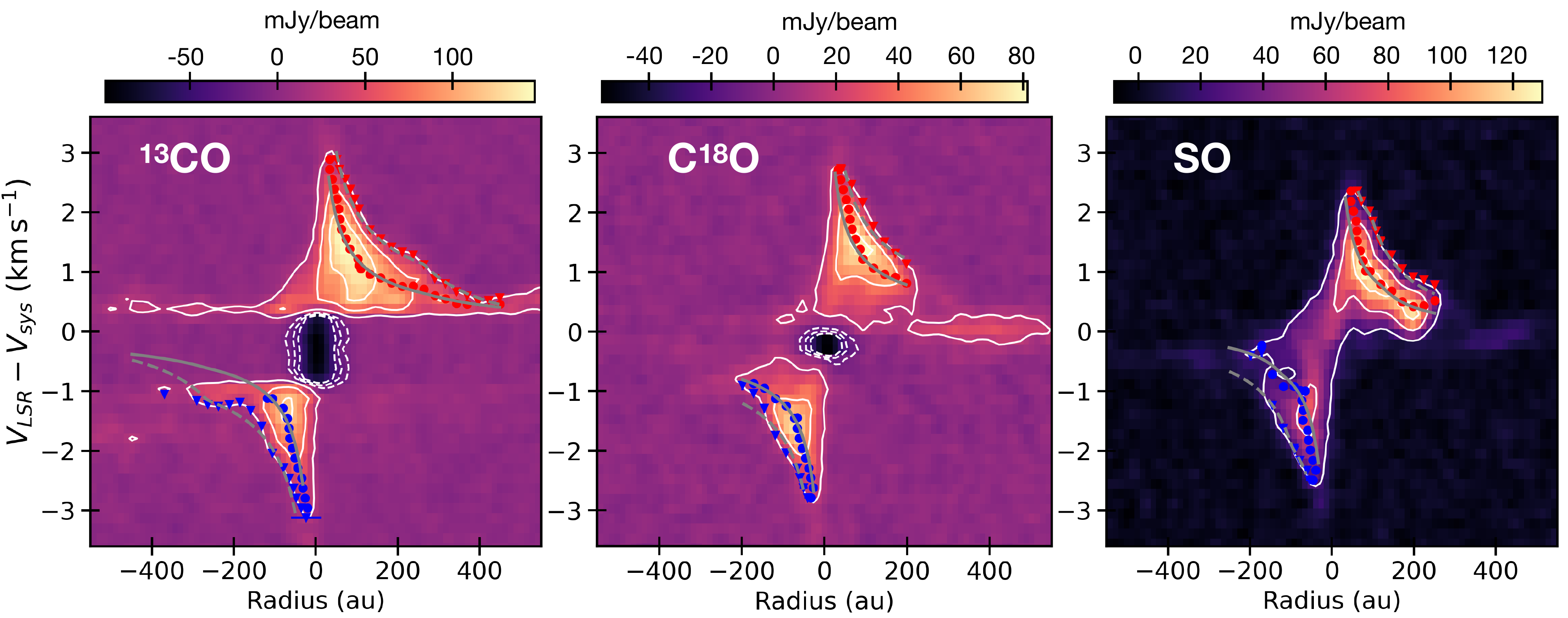}
\caption{The left, middle, and right panels show PV diagrams along the major axis for the $^{13}$CO, C$^{18}$O, and SO molecules, respectively. A 5-pixel average ($\sim0\farcs5$) PV profile was obtained in all the figures. The white contours mark the emission at 5$\sigma$, 15$\sigma$, and 25$\sigma$ to match the contours shown in their channel maps (Figures \ref{fig:channel_maps_13CO}, \ref{fig:channel_maps_C18O_large}, and \ref{fig:channel_maps_SO} in the Appendix). The dashed white contours in the CO molecules show the negative emission at -3$\sigma$, -6$\sigma$, and -9$\sigma$ caused by continuum over subtraction. The ridge and edge methods to trace the disk emission are marked by filled circles and triangles. The results from SLAM indicate that the inner portions of the velocity profiles can be reproduced by a central protostellar mass of $\sim0.5 \; \rm M_\odot$.}
\label{fig:CO_PV_digrams_SLAM}
\end{figure*}

First, we obtained PV cuts along the major axis of the disk for the $^{13}$CO and C$^{18}$O isotopologues (5-pixel wide or $\sim 0\farcs5$) and show them in Figure \ref{fig:CO_PV_digrams_SLAM}. In both PV, we observe differential rotation, emission asymmetries caused by envelope contribution, and interferometric filtering. The most prominent signatures of infalling motion can be seen in the $^{13}$CO PV diagram, as it displays a characteristic emission excess in the top-left non-Keplerian quadrant (quadrant II). This emission has been described in several works, using simple infall models \citep[e.g.,][]{vantHoff2018, Momose1998}. 


\begin{deluxetable*}{lccccc}
\tablewidth{20pt}
\tablecolumns{5}
\tablecaption{Best fit parameters from SLAM. \label{table:slam_parameters}}
\tablehead{
\colhead{Line (method)} & \colhead{$p_{in}$} & \colhead{$p_{in}+dp$} & \colhead{$R_b$ (au)} & \colhead{$M_\star$ ($M_\odot$)} & \colhead{$V_{sys}$ (km/s)} }
\startdata
\hline
$^{13}$CO (edge) & $0.59 \pm 0.01$ & $1.6 \pm 0.1$ &$264 \pm 3$ & $0.73 \pm 0.02$ & $2.80 \pm 0.01$  \\
$^{13}$CO (ridge) & $0.67 \pm 0.02$ & $0.9 \pm 0.1$ &$265 \pm 22$ & $0.22 \pm 0.03$ & $2.82 \pm 0.01$  \\
\hline
C$^{18}$O (edge) & $0.58 \pm 0.01$ & -& - & $0.66 \pm 0.01$ &  $2.82 \pm 0.01$ \\
C$^{18}$O (ridge) & $0.60 \pm 0.01$ & - & - & $0.33 \pm 0.01$ &  $2.81 \pm 0.1$ \\
\hline
SO (edge) & $0.48 \pm 0.19$ & $1.13 \pm 0.23$ & $97 \pm 12$  & $0.76 \pm 0.19$ &  $3.05 \pm 0.02$ \\
SO (ridge) & $0.83 \pm 0.06$ & $1.30 \pm 0.26$ & $120 \pm 39$ & $0.14 \pm 0.14$ &  $3.08 \pm 0.01$ \\
\enddata
\tablecomments{The indicated uncertainties are the 14th and 68th statistical uncertainties from the MCMC analysis from SLAM. The adopted disk and stellar properties are obtained by averaging the results from the edge and ridge methods.}
\end{deluxetable*}

We fit the full functional form of equation (\ref{eq:slam}), which consist of two  power laws, to the $^{13}$CO emission. We did this because we observe emission above $5\sigma$ extending to distances $>200$~au along the major axis. By using equation (\ref{eq:slam}), we were able to derive the change in the power law index in the outer parts $dp$ as well as the braking radius $R_b$. For C$^{18}$O, we only fit a single  power law (the first portion of equation \ref{eq:slam}) as the extended emission along the major axis is weaker. In both cases, we assumed a distance of $d$=132~pc, a disk inclination of $i=47\degree$, and excluded the filtered-out velocity channels from the fit. In Figure \ref{fig:CO_PV_digrams_SLAM}, we overlaid the best-fit model from SLAM to the PV diagrams along the major axis of $^{13}$CO and C$^{18}$O. The two sets of lines (interior and exterior) in quadrants I and III represent two fitting methods of SLAM: the ridge (circles) and the edge (triangles) modes, respectively.

For both CO isotopologues, we find almost Keplerian inner  power law indices of $p_{in}\sim0.6\pm 0.04$ using the edge and ridge methods (second column in Table \ref{table:slam_parameters}). We take the median value of the different methods as the most likely inner  power law index. The stellar mass obtained from SLAM relates to this inner  power law index, and by forcing Keplerian rotation, we derive a mass of $M_\star = 0.5 \pm 0.2$~$M_\odot$ (fifth column in Table \ref{table:slam_parameters}). Our derived dynamical mass, is approximately consistent with the one derived by \cite{Brinch2013} of $\rm M_\star =0.4 \, M_\odot$ after re-scaling by our chosen source distance ($d$) and disk inclination angle ($M_\star \propto \frac{d}{\sin^2({i_{disk}})}$).

We derive the radius of the rotating gas disk from the fit to the $^{13}$CO emission, and the radius is defined as the point where the inner and outer  power law change in functional form ($\rm R_b$), i.e., where the quasi-Keplerian disk meets the envelope emission. We obtain that such a radius is at $R_b = 265 \pm 9$~au using ridge and edge methods. The calculation of the difference of the  power law index from the inner to the outer parts $dp$ yields significantly different values between the edge and ridge methods. The $p_{in}+dp$ values measured for the edge and ridge method are 1.6 and 0.9, respectively, which gives an average of $p_{in}+dp=1.2 \pm 0.4$, where the uncertainty is dominated by the standard deviation. A $p_{in}+dp \sim 1$ means that the  $^{13}$CO emission at $>260$~au ($>2\arcsec$) follows approximately an infalling velocity profile.

\subsubsection{SO emission}
We also use SLAM to understand the kinematics of the SO emission along the major axis of the disk (right panel of Figure \ref{fig:CO_PV_digrams_SLAM}). The edge method shows that the SO material is rotating at Keplerian speed ($p_{in}\sim 0.5$) up to a radius of $R_b\sim$100~au, while the ridge method shows that the SO emission has an inner  power law of $p_{in} \sim 0.8$ up to $\sim$100~au, indicating a value in-between Keplerian and pure infall. Both methods fit the velocity of the outer region ($R\gtrsim100$~au) with a  power law broadly consistent with infall $p_{in}+dp \sim 1.0$. This results suggest that the SO spirals, which are located beyond $\sim$100~au, rotate slower than Keplerian and are likely infalling. The results from SLAM are consistent with what is seen in the SO moment 1 map in Figure \ref{fig:Disk-size-comparison}, where the inner $\sim$1$\arcsec$ ($\sim$130 au) of the emission shows a strong velocity gradient along the disk direction while the outer part has a sudden change in the direction of the velocity gradient, indicating infall. Since CO and SO likely trace different physical conditions in the disk and envelope, such as different layers or altitudes from the observer's point of view, it is not surprising that the Keplerian radii of the tracers are different.


In summary, the kinematic analysis shows us that 1) the compact gas structure identified in previous sections has an almost Keplerian rotation profile extending to $\sim$260~au in CO and $\sim$100~au in SO, 2) there is evidence of infalling envelope beyond the quasi-Keplerian velocity profile in the CO and SO lines, and 3) the mass derived for the central (proto)stellar source is approximately $ \rm M_\star = 0.5 \pm 0.2 \, M_{\odot}$.

\begin{figure*}[ht!]
\centering
\epsscale{1.1}
\plotone{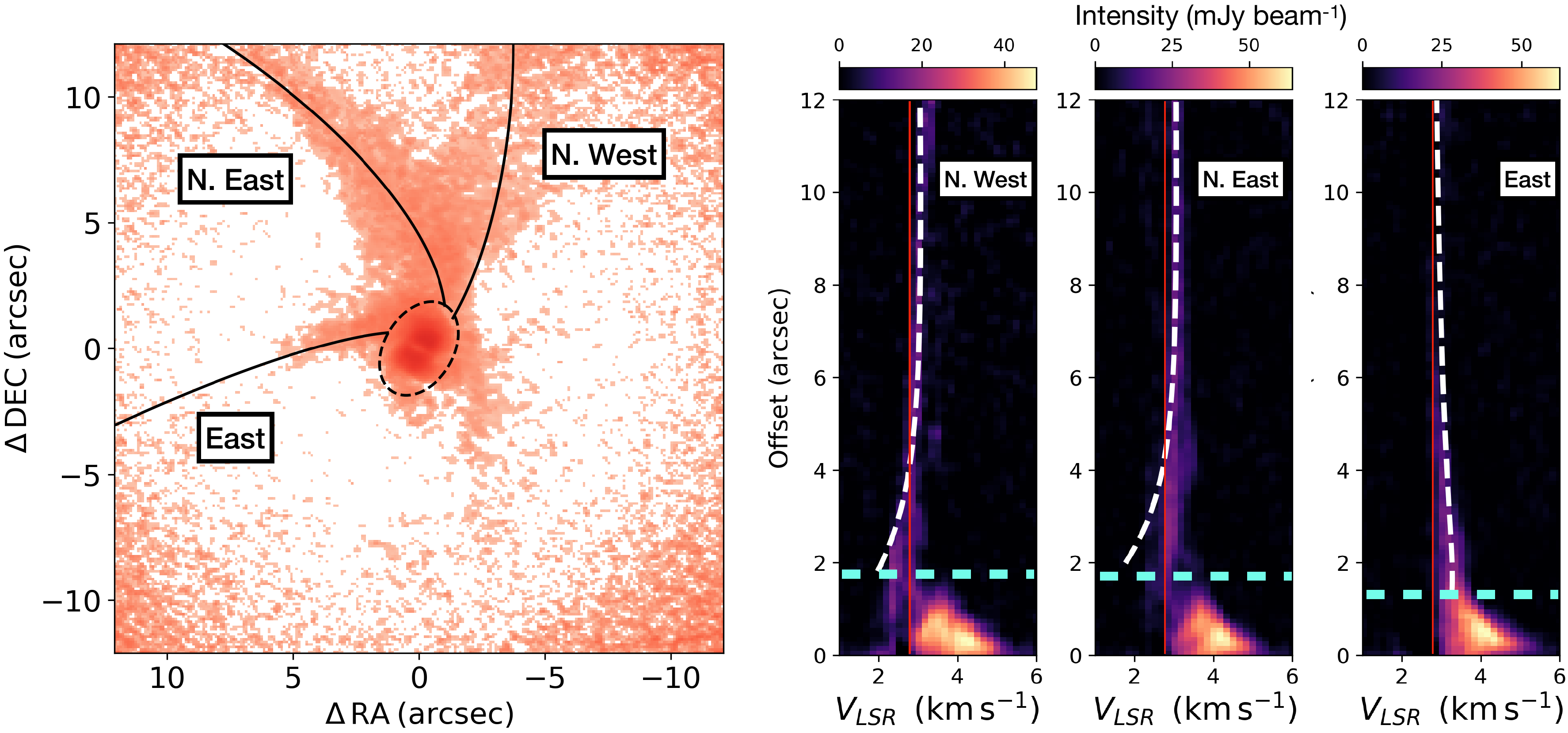}
\caption{Left: A streamer model overlaid on the moment 0 map of the C$^{18}$O emission. The model was used to reproduce three features: the conspicuous northern streamer that divides into a northeast and a northwest branch and the short eastern arm. The dashed oval corresponds to the size of the quasi-Keplerian radius of 265~au as calculated from SLAM. Right: The projected distance along the streamer vs. velocity is shown for the northwest, northeast, and east streamers, as labeled. The white dashed lines show the spatial and velocity trails of the models that delineate the observed position-velocity diagrams. The horizontal cyan lines mark the distance at which the infalling streamers transition to a mostly Keplerian velocity. The thin red lines mark the systemic velocity at $v_{sys} = 2.8 \, km s^{-1}$.  
}
\label{fig:C18O_streamer_model}
\end{figure*}

\subsection{Infalling streamer model}
\label{subsec:infall_streamer_model}

Two molecular tracers presented in Section \ref{subsec:results_Spirals} exhibit a spiral or arc-like structure. C$^{18}$O displays a large-scale arc structure detected at the northern edge of the map, which connects to the disk on the north and northwest sides. There is also a smaller-scale  C$^{18}$O arm towards the east of the disk (see Figure \ref{fig:C18O_map}). 
The SO emission, on the other hand, exhibits three spiral-like structures. As discussed later in Section \ref{sec:disc_SO_spirals}, these spiral features might reflect temperature enhancements possibly induced by the impact of the ambient cloud materials on the disk surface. Thus, we would like to test whether their morphology can be reproduced by a collapsing rotating cloud that accretes to a central gravitational object. For this, we follow the analytical solutions proposed by \cite{Mendoza2009}. Similar streamer modeling work for other sources has been presented in e.g., \cite{Yen2017, Yen2019}, \cite{Pineda2020}, \cite{Thieme2022}, \cite{Valdivia-Mena2022}, and \cite{Kido2023}.

In brief, the \cite{Mendoza2009} model is described by a gravitational potential produced by the central star and a rotating gas cloud with a finite radius $r_0$, where particles can collapse starting with a radial velocity $v_0$. We use spherical coordinates to trace the motion of the infalling particles, where $r$ is the radial coordinate, $\theta$ is the polar angle (where the zenith is defined by the disk's rotational axis), and $\phi$ is the azimuthal angle (which moves in the plane of the disk) \citep[see Figure 1 in][]{Mendoza2004}. For simplicity, we assume that the rotational axis of the cloud and the disk are aligned ($i=47\degree$, P.A.= $149\degree$, the near side is $59\degree$ east of north). 

The \cite{Mendoza2009} model relaxes the assumption of the classical \cite{Ulrich1976} model, and it differentiates as the former assumes that the cloud has a finite size and that accreting particles might start its collapse with a non-zero initial radial velocity. These differences translate into trajectories described by conical sections rather than parabolic motions. The assumptions involved in this model are: 1) The pressure gradient of the flow is negligible, 2) the particles at the boundary of the spherical cloud rotate as a rigid body, 3) self-gravity is negligible, and thus the massive central source dominates the motion. The first assumption is valid if the flow is supersonic and the viscosity and radiation effects are small enough. The second one is necessary to give the material a distribution of angular momentum. The final requirement is not difficult to overcome, considering that the central protostar is more massive than the envelope and disk.

To model C$^{18}$O, we defined the inner edge of the cloud to be of size $R_{cloud} \sim 5300$~au (20 times the disk's radius), fixed the mass of the protostar to $M_{\star} =0.5 \, M_\odot$ protostar, and set the initial angular specific momentum of the particles to $h_0 = \sqrt{R_{disk} \, GM_\star} \sim 1250$~$\rm au \, km \, s ^{-1}$, where $G$ is the gravitational constant, and $R_{disk} = 265$~au is the size of the disk. For this molecule, we named the three most prominent arc-like features as East, N. East and N. West, as shown on the right panel of Figure \ref{fig:C18O_streamer_model}. We manually adjusted the $\theta$ and $\phi$ angles to match the  position and velocities of the three arc-like features. As shown in Figure \ref{fig:C18O_streamer_model}, the streamer model reproduces well the projected positions as well as the velocity structure along the trajectories of the C$^{18}$O line.  

\begin{figure*}[ht!]
\centering
\epsscale{1.05}
\plotone{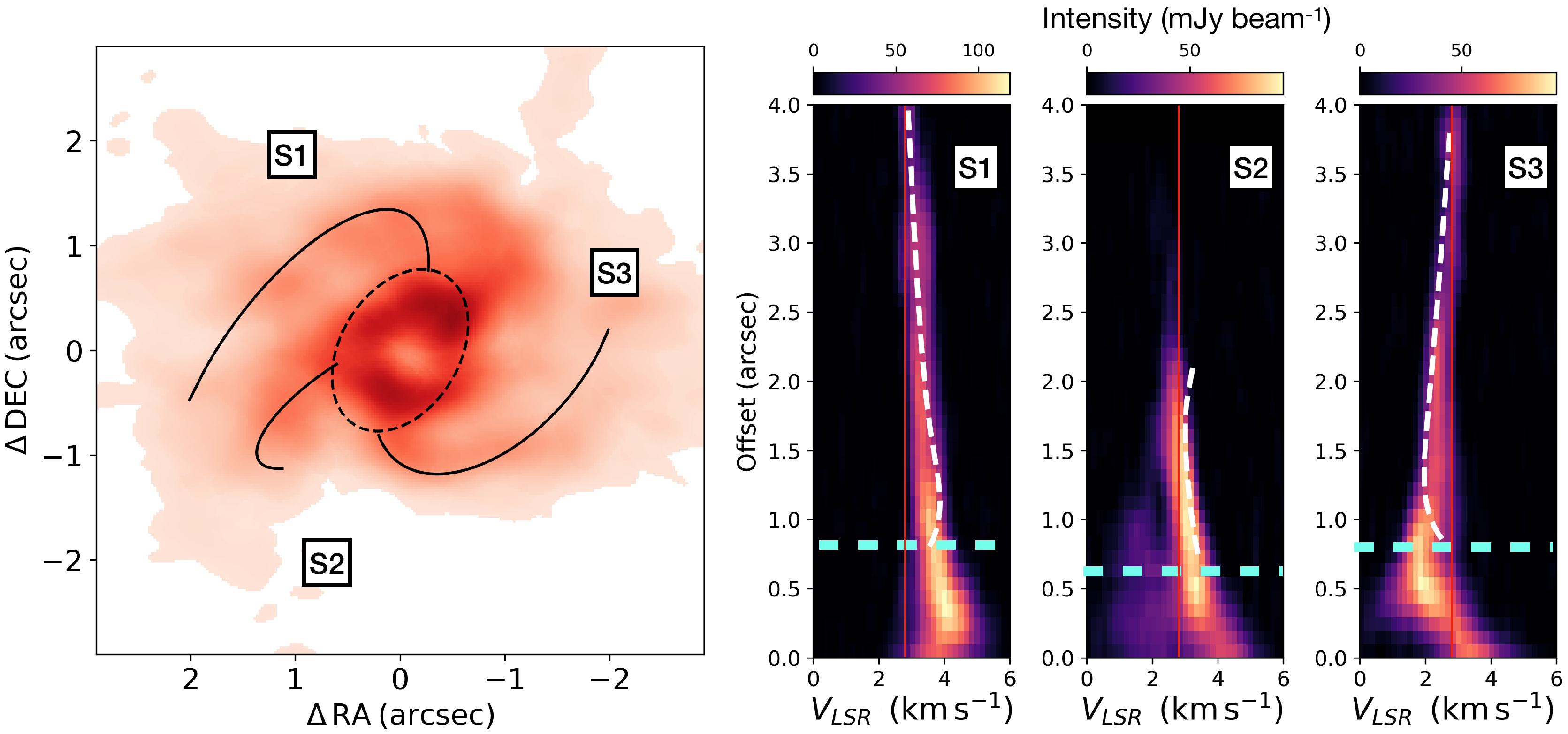}
\caption{Left: A streamer model overlaid on the moment 0 map of the SO emission. We attempted to reproduce the three SO spirals. The dashed oval corresponds to the size of the Keplerian radius of 110~au as calculated from SLAM. Right: The projected distance along the streamer vs. velocity is shown for spirals, as labeled. The white dashed lines show the spatial and velocity trails of the models that delineate the observed position-velocity diagrams. The horizontal cyan lines mark the distance at which the infalling streamers should be affected by Keplerian rotation. The thin red lines mark the systemic velocity at $v_{sys} = 2.8 \, km s^{-1}$.  
}
\label{fig:SO_streamer_model}
\end{figure*}

For SO, we set a smaller disk radius of $R_{disk} = 110$~au (following the results from SLAM on SO), and to better match the tightness of the spirals, we had to define a much smaller initial position of the streaming particles of only 275 au to 1320 au to match different spirals. Then, we varied $\theta$ and $\phi$ to visually match the three spirals features detected in SO. In this way, we were able to reproduce the spatial characteristics of the SO spirals as shown in Figure \ref{fig:SO_streamer_model}. The velocity profiles of the three streamer (white dashed lines) follow well the emission seen in the PV diagrams, meaning that their velocity structures of the SO spirals can be reproduced by infalling streamer models (right panel in Figure \ref{fig:SO_streamer_model}). 

For both C$^{18}$O and SO lines, we marked the radial position (with a cyan horizontal line) where the infalling profile should turn into a Keplerian dominated profile as shown in our SLAM analysis (Section \ref{subsec:SLAM_analysis}). Physically this means that the infalling streamer do not reach the midplane of the disk, but instead they likely landed at the disk's surface at distances of several tens of au above the midplane level.

We acknowledge that since we have not performed a mathematical fitting procedure, the values shown in Table \ref{table:streamer_c18o_parameters} might not be the best or even unique solutions to match the observations. In fact, for each model trajectory, small changes in the $\theta$ and $\phi$ angles also produce similarly valid results. These small ranges of solutions could, for example, represent that these streamers have an intrinsic 3-dimensional width.

\begin{deluxetable}{lcccc}
\tablewidth{20pt}
\tablecolumns{4}
\tablecaption{C$^{18}$O and SO streamers parameters used in Figures \ref{fig:C18O_streamer_model} and \ref{fig:SO_streamer_model}. \label{table:streamer_c18o_parameters}}
\tablehead{
\colhead{Streamer} & \colhead{$\theta$} & \colhead{$\phi$} & \colhead{$R_{cloud}$}\\
\colhead{ } & \colhead{degrees ($^\circ$)} & \colhead{degrees ($^\circ$)} & \colhead{au}
}
\startdata
\hline
C$^{18}$O - East & 57 & 188 & 5300  \\
C$^{18}$O - N. East & 85 & 270 & 5300 \\
C$^{18}$O - N. West & 108 & 300 & 5300 \\
\hline
SO - S1 & 124 & 260 & 550  \\
SO - S2 & 55 & 155 & 275 \\
SO - S3 & 67 & 75 & 1320 \\
\enddata
\tablecomments{$\theta$ values smaller than $90^\circ$ correspond to streamers landing on the far face of the disk (aligned with red-shifted outflow). $\phi$ values increase counter clock-wise, and $\phi= 0^\circ$ is defined at P.A.= $329^\circ$, i.e., $180^\circ$ from the disk position angle of $149^\circ$.}  
\end{deluxetable}

\section{Discussion} 
\label{sec:Discussion}

\subsection{The SO ring and spirals}
\label{sec:disc_SO_spirals}

We have previously mentioned that we were unable to confirm whether the ring-like structure is a real ring or the product of continuum over-subtraction. These claims come mainly from a radiative transfer analysis of a three-layer model of dust and gas, where the dust continuum is concentrated in the disk mid-plane in the form of  a thin layer as depicted in Figure 11 of \cite{Yen2019}. In such a case, radiative transfer equation can be written as:

\begin{align} 
    I_{\nu} =& [B_{\nu}(T_{g}) (1+ e^{-\tau_d - \tau_g}) - B_\nu(T_d)(1-e^{-\tau_d})] \\
    &(1-e^{-\tau_g}) \nonumber
\end{align}


Where the subscripts $d$ and $g$ stand for dust and gas, respectively, $\tau$ is the optical depth, and $B_\nu$ is the Plank function at temperature $\rm T$. Additionally, when the dust is optically thick ($\tau_d \gg 1$), as near the stellar position (where $\rm T_{d}\sim 170$~K), it can be written as:

\begin{equation}
\label{eq:radiative_transfer_eq}
    I_{\nu} = [B_{\nu}(T_{g}) - B_{\nu}(T_{d})] (1- e^{- \tau_g})
\end{equation}

A natural conclusion from equation (\ref{eq:radiative_transfer_eq}) is that the temperature of SO must be higher than the temperature of the dust everywhere in the disk, as negative SO emission is not observed in any channel map ($\rm I_{\nu}>0$). Then, two scenarios are possible to
to explain the lower SO emission at the disk center.  We could have a real ring structure, which would imply a very small SO optical depth towards the center $\tau_g$ (i.e., a small (1- $e^{-\tau_g}$) factor). Alternatively, we could have abundant SO gas in the center, i.e. (1- $e^{-\tau_g}$)$\sim 1$, but the apparent emission drop happens due to SO being marginally hotter than the dust $\rm T_{g} \gtrsim T_{d}$. This second scenario, would be consistent with the continuum subtraction case, i.e., there is no real ring.  Since we cannot distinguish between these two possibilities, we will refer to this feature as a ring. However, we caution the reader that  more observations are needed to confirm that this is not a purely continuum-subtraction artifact.

An additional support to the ring being a real structure comes from the off-centering between the continuum emission and SO emission depression. We speculate that if the dust were the cause of a central SO continuum over-subtraction, then the SO emission depression should coincide with the dust maximum emission, which is not the case (see Sec. \ref{subsec:results_Spirals}).

An SO ring tracing heated material has been proposed to explain ALMA observations of the Class I protostar L1489 IRS \citep{Yen2014,Yamato2023}, the Class 0 protostar IRAS 16253-2429 \citep{Aso2023}, and the embedded source  L1527 IRS (IRAS 04368+2557) \citep{Ohashi2014,Sakai2014}. While these structures are often inferred from PV diagram analysis, a bright ring-like structure in SO can be directly observed in the disk around Oph IRS63 due to its moderate inclination. The observed SO enhancement, could result from OH reactions after the liberation of S via sputtering \citep{Hartquist1980}, or SO could result directly via sublimation from dust grains when temperatures are $\gtrsim40-60$~K \citep[e.g.,][]{Hasegawa1993}. In either case, the temperature increase needed to enhance SO might have originated at the landing zone of the infalling material, i.e., the SO ring could trace an impact zone of the infalling envelope on the disk, as predicted, for example, in the slow shock models of \cite{Aota2015}. This scenario would naturally connect the origin of the SO ring with the observed spirals structures. Further evidence of this connection comes from the fact that the outer edge of the SO ring approximately marks a transition between the infalling spirals structures and the inner $100$~au that rotate closer to Keplerian velocity. Such kinematic connection has been proposed, for example, for L1527 IRS through the centrifugal radius, where a large portion of the kinetic energy of the infalling material is transformed to rotational energy allowing the gas to orbit the central source \citep[e.g.,][]{Sakai2014}.

Furthermore, some of the features observed in SO have been predicted through  magneto-hydrodynamical (MHD) models. Infalling and rotating spiral structures have beeen reported when there is a misalignment between the mean magnetic field and rotation axis of a collapsing core \citep[e.g.,][]{Li2013, Vaisala2019, Wang2022, Tu2023}. These MHD simulations have even found a connection between shock regions and spirals of streaming material, as well as formation of ring-like structures due to the circularization of spiraling inflows of matter \citep{Wang2022}.

\subsection{Mass infalling rate}

Since an infalling envelope model reproduces the arc-like structures of C$^{18}$O, as well as, the spirals in SO, we would like to calculate how much mass these streamers transfer from the inner part of the envelope to the disk. To estimate this, we use the C$^{18}$O (2-1) emission as it is optically thin, and thus we can calculate the amount of mass traced by it. We first confirm that C$^{18}$O (2-1) is optically thin using the results obtained by \cite{vantHoff2018}. They found that if the ratio of the brightness temperature of $^{13}$CO (2-1) to C$^{18}$O (2-1) is higher than 1.5, then C$^{18}$O (2-1) emission is optically thin. These results were tested using a range of gas temperatures (20~K to 100~K) and column densities ($10^{13}$~cm$^{-2}$ to 10$^{18}$~cm$^{-2}$) assuming standard isotope ratios of [$^{12}$C]/[$^{13}$C] = 77 and [$^{16}$O]/[$^{18}$O] = 560 \citep{Wilson1994}. For the gas around Oph~IRS63, we find that the ratio between the peak brightness temperature of the $^{13}$CO and C$^{18}$O lines is between 2 and 3 in the region where most of the CO gas is detected. This result confirms that C$^{18}$O is optically thin, and because the ratio is $<4$, it also confirms that $^{13}$CO is optically thick \citep[see][]{vantHoff2018}. 

Since $^{13}$CO is optically thick,  we can use it to calculate the excitation temperature of the gas according to equation (\ref{eq:excitation_temp}). The excitation temperature ($T_{ex}$) coincides with the kinetic temperature when the gas is in Local Thermodynamic Equilibrium (LTE). In this equation, $h$, $\nu$, and $k$ refer to Plank's constant, the frequency of the emission, and the Boltzmann constant, respectively. $T^{13}$ is the peak of the brightness temperature inferred from the $^{13}$CO (2-1) line, and $J(T_{bg})=\frac{h \nu / k}{ e^{h \nu / k T_{bg}}- 1}$ is the intensity expressed in units of temperature for the microwave background temperature of 2.7~K.

\begin{equation}
    T_{ex} = \frac{h\nu/k}{\ln(1+ \frac{h\nu/k}{T^{13}+J(T_{bg})})}
    \label{eq:excitation_temp}
\end{equation}

From this equation, we measure a peak excitation temperature of 21~K in the envelope. Once the excitation temperature is obtained, we then use the C$^{18}$O emission to calculate the gas mass of the streamer according to equations (\ref{eq:column_density}) and (\ref{eq:mass}) \citep[following e.g.,][]{Zapata2014,Lopez-vazquez2020}. In these equations N(J) is the column density of the J level obtained from Boltzmann's equation, $B_e$ is the rotational constant of the molecule, Z is the partition function of C$^{18}$O, $m(H_2)$ is the molecular hydrogen mass, and $\Delta\Omega$  is the angular size of the emission, and $d$ is the distance to Oph IRS63.
    
\begin{eqnarray}
    N_{total} =& N(J) \frac{Z}{(2J+1)}  e^{ \frac{h B_e J(J+1)}{k T_{ex}}} \label{eq:column_density} \\ 
    M(H_2) =& m(H_2) \frac{X(H_2)}{X(C^{18}O)} \Delta \Omega d^2 N_{total}
    \label{eq:mass}
\end{eqnarray}

We consider a region that encompasses the extended C$^{18}$O emission but excludes the disk's emission, which is concentrated in the inner 265~au.  Using a standard abundance of $3 \times 10^{-7}$ \citep{Frerking1982} for C$^{18}$O and $B_e$ value taken from the JPL database \citep{Pickett1998}, we obtain a total C$^{18}$O mass of $M_{envelope} \sim 0.04~M_\odot$. The main uncertainty in obtaining this mass comes from the assumed abundance of C$^{18}$O with respect to H$_2$. Additionally, our mass estimate is likely a lower limit value due to the flux missed by the interferometer. This is already suggested by the fact that our value is about half of the total envelope mass derived from \cite{Brinch2013}, which was 0.07$~M\odot$.

Finally, we estimate a mass infalling rate from the envelope to the disk of $\dot{M}_{infall} \sim 4\times 10^{-6}$~$M_\odot \,  yr^{-1}$. To do this, we divided the mass contained in the infalling envelope by the infalling time scale ($\dot{M}_{infall}\sim M_{envelope}/t_{infall}$), where $t_{infall}$ is the time it takes a parcels of gas at the outermost edge of the C$^{18}$O emission to reach the disk. As mentioned in previous works that calculate such mass infall rates \citep[e.g.,][]{Yen2019, Ohashi1997,Takakuwa2013}, this quantity is just an order of magnitude estimate, it could be higher if we consider that part of the large-scale emission can be filtered out by the interferometer, or it could be lower if the infalling velocity is slower than free-fall.

\subsection{(proto)Stellar mass accretion rate}

\begin{figure}[ht!]
\centering
\epsscale{1.1}
\plotone{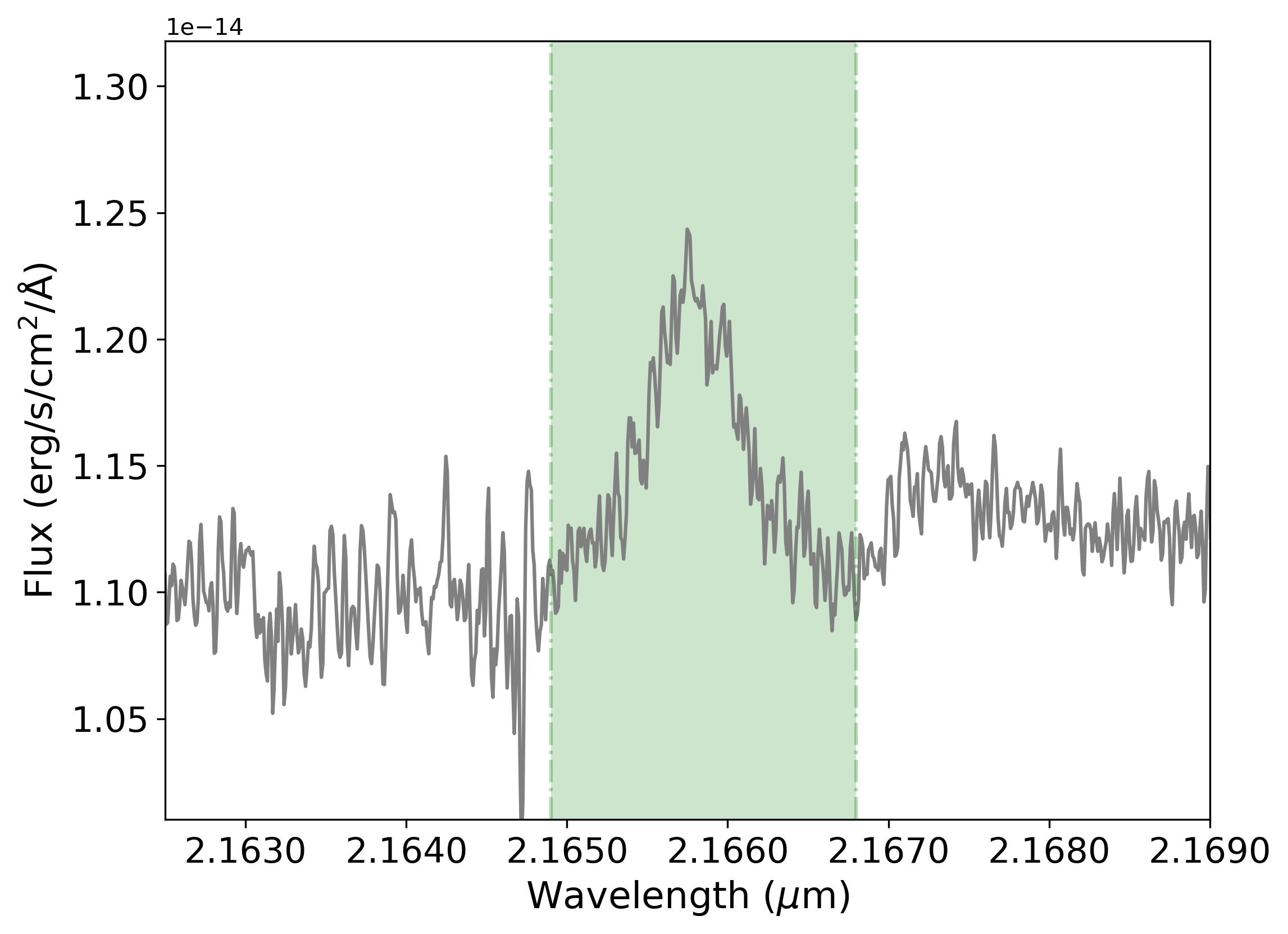}
\caption{iSHELL K-band spectrum of Oph IRS63 centered around the Br$\gamma$ line at $2.166\, \micron$. The green-shaded region depicts the region integrated to obtain (proto)stellar mass accretion rate. This spectrum shows the flux before extinction correction. }
\label{fig:iSHELL_Br_gamma}
\end{figure}

As with most young stellar sources, we expect some material to be funneled from the inner parts of the disk to the central (proto)star. This mass transfer is typically traced via Hydrogen recombination lines, such as H$\alpha$ in the optical, Pa$\beta$ or Br$\gamma$ in the near-infrared \cite[e.g.,][]{Muzerolle1998,Hartmann2016}.  Due to the embedded nature of Oph IRS63, optical tracers cannot be used to measure the mass accretion rate. Thus, we use the Br$\gamma$ line at 2.166~$\micron$ obtained from the SpeX and iSHELL near-infrared spectra (see Appendix \ref{appendix:infrared}). These spectra are pseudo-flux calibrated, and a region around the 2.166~$\micron$ Br$\gamma$ line of the iSHELL spectrum is presented in Figure \ref{fig:iSHELL_Br_gamma}. 

We deredden the infrared spectrum using our $A_v=24$~mag derived from the procedure described in Appendix \ref{appendix:infrared}. We integrate the Br$\gamma$ emission in the spectra to calculate the line flux ($F_{line}$) and the line luminosity ($L_{line}$) using equation (\ref{eq:line_lum}).

\begin{eqnarray}
    L_{line} &=& 4\pi d^2 F_{line} \label{eq:line_lum}\\ 
    \log{ \left( \frac{L_{acc}}{L_{\odot}} \right)} &=& a \log{ \left( \frac{L_{line}}{L_{\odot}} \right)} + b \label{eq:acc_luminosity} \\
    \dot{M}_{acc} &\sim& \left( 1-\frac{R_{\star}}{R_{in}} \right) \frac{L_{acc} R_{\star}}{ G M_{\star}} \label{eq:stellar_accretion}
\end{eqnarray}

We convert the line luminosity to accretion luminosity following the prescription in \cite{Alcala2017} (equation \ref{eq:acc_luminosity}), where $a=1.19 \pm 0.1$ and $b=4.02 \pm 0.51$ for the Br$\gamma$ line. Finally, we transform this line measurement into a stellar mass accretion rate via equation (\ref{eq:stellar_accretion}), assuming $R_{in}=5R_{\star}$ \citep{Hartmann1989} and $R_{\star} = 3.0 \, R_\odot$ (this assumes a gravity of $\log{g}=3.2$, which is consistent with values found around Class0/I protostars (\citealt{Doppmann2005,Greene2018}, and Flores et al. (submitted)). Putting all these numbers together, we measure a (proto)stellar accretion luminosity of $L_{acc} = 0.21$~$L_\odot$, and consequently, a mass accretion rate of $\dot{M}_{acc} = 5\times 10^{-8}$~$M_\odot \, yr^{-1}$, for an $Av$ of 24~mag. To estimate the uncertainties of this stellar accretion measurement, we varied the $Av$ from 20 to 40 mag of extinction, and the stellar radius between 2.0 and 3.7 $R_\odot$ (i.e., $ \log{g}= 3.0 - 3.5$) and obtain (proto)stellar mass accretion rates from $\dot{M}_{acc} = 2\times 10^{-7}$ to $2\times 10^{-8}$~$M_\odot \, yr^{-1}$. Although this method has traditionally been applied to T Tauri stars or Class II sources \citep{Alcala2017,Manara2019b}, it has recently been demonstrated to also work for Class I and Flat Spectrum sources  \citep[e.g.,][]{Fiorellino2021,Fiorellino2022}.

An independent check on the mass accretion rate for Oph~IRS63 can be obtained following the standard assumption that the bolometric luminosity is the sum of the stellar and the accretion components (equation \ref{eq:Lbol}).

\begin{equation}
\label{eq:Lbol}
    L_{bol} = L_\star + L_{acc}
\end{equation}

For Oph~IRS63, the bolometric luminosity $L_{bol}$ is 1.3~$\rm L_\odot$ \citep{Ohashi2023}, while the stellar luminosity ($\rm L_\star$) using $R_\star=3.0 \, R_\odot$ and $T_{eff} = 3500$~K can be calculated as
\begin{equation}
L_\star = 4\pi \, \sigma \, R_\star^2 \, T^4_{eff} = 1.21~\rm L_\odot
\end{equation}

This means that the accretion luminosity is of the order of $L_{acc} \sim 0.1 \, L_\odot$, which corresponds to a mass accretion rate of $\dot{M}_{acc} \sim 3 \times 10^{-8}$~$M_\odot \, yr^{-1}$ according to equation (\ref{eq:stellar_accretion}). This value agrees with the lower range of the mass accretion rate derived from the Br$\gamma$ line. 

\subsection{Envelope infall vs. stellar accretion rate}
\label{subsec:envelope_infall_Toomre-Q}
We observe a significant discrepancy between the mass infall rate $\dot{M}_{infall} \sim 4\times 10^{-6}$~$M_\odot \,  yr^{-1}$  and the stellar accretion rate $\dot{M}_{acc} = 2-20 \times 10^{-8}$~$M_\odot \,  yr^{-1}$, i.e., between the mass infalling from the envelope to the disk and the mass transfer from the disk to the (proto)star. 

If these measurements represent their respective ``accretion process" and are not just transients, then the natural consequence is a mass build-up in the disk. This is because more mass is infalling into the disk than leaving it, assuming the mass outflow rate is  $\sim 10\%$ of the mass accretion rate \citep{Ellerbroek2013}. 

During the embedded phase, envelope accretion and viscous processes naturally increase a disk's mass and size as the  angular momentum is redistributed. However, when a disk becomes too massive, it can become gravitationally unstable forming asymmetries and spiral features \citep[e.g.,][]{Kuffmeier2018} and sometimes fragmenting \citep[e.g.,][]{Vorobyov2006,Das2022}. The Toomre-Q parameter \citep{Toomre1964} is a standard metric used to determine whether a disk is gravitationally stable ($Q\gtrsim1$) or unstable ($Q\lesssim 1$). We calculate the Toomre-Q parameter for the disk around Oph~IRS63 using equation (\ref{equation:toomre}).

\begin{equation}
    Q \equiv \frac{c_{s} \Omega}{ \pi G \Sigma} \label{equation:toomre}
\end{equation}

We obtain the angular speed of the gas disk from $\Omega= \sqrt{GM_\star/R_{disk}^3}$ and approximate the disk surface density by $\Sigma \sim M_{disk} / \pi R^2_{disk}$. The gas sound speed ($c_s$) is calculated using $c_s = \sqrt{k_B T /\mu m_{\rm H}} $, where $k_B$ is Boltzmann's constant, $m_{\rm H}$ is the mass of a hydrogen atom, and $\mu$ is the mean molecular weight that is often taken to be $\mu=2.3$ in disks \citep[e.g.,][]{Armitage2015}. As pointed out in Section \ref{subsec:continuum}, we can use two disk temperatures: $T=20$~K or $T=46$~K, which yield different dust disk masses. We calculate  the total gas disk mass assuming a standard gas-to-dust ratio of $g/d=100$ and obtain $M_{disk}\sim0.05 \, M_\odot$ and $M_{disk}\sim0.02 \, M_\odot$, respectively. We obtain a Toomre-Q parameter ranging from $Q \sim 2-8$ from these two temperatures.  It is worth mentioning that the main uncertainties in deriving these values are that the gas-to-dust ratio could be different from 100 and the dust mass is likely a lower-limit.

Although it is not possible to predict whether these infall and mass accretion rates will remain 
at the current levels, it can be calculated that it would take approximately $\sim 10^{4} \, yr$ for the disk to achieve a mass of $M_{disk}\sim0.1 \, M_\odot$. If this happens, the Toomre-Q parameter would be close to unity, making the disk gravitationally unstable, and possible triggering episodic accretion bursts.

\subsection{Gas vs. dust disk size}

Comparing the extension of gas and dust in protoplanetary disks is central to the question of radial drift \citep{Facchini2019,Trapman2020} and dust settling \citep{Villenave2020}, and ultimately to the planet formation process itself \citep{Armitage2015}. Measuring gas and dust radial sizes can inform us whether grain growth has begun and how dynamically coupled these two components are. 

In Section \ref{subsec:continuum}, we measured the dust disk size as the radius enclosing 95$\%$ of the flux and found $R_{95\%} = 0\farcs47$ or $R_{95\%} = 62$~au (at $d=132$~pc). Then, in Section \ref{subsec:SLAM_analysis}, we use the kinematics of $^{13}$CO and C$^{18}$O to determine that the disk around Oph~IRS63 follows an almost Keplerian rotation up to $265\pm9$~au. By dividing the gas and dust disk sizes, we find a gas-to-dust size ratio of $R_{gas}/R^{dust}_{95\%}\sim4.3$. This gas-to-dust size ratio might not be directly comparable to studies in Class II disks because, in such cases, both the dust and gas disks' radii are measured based on a fixed amount of flux enclosed in a certain radius (e.g., 90$\%$ or 95$\%$). However, due to the embedded nature of our source, we could not perform such a measurement for the gas and instead used a different definition than for the dust. In any case, for the sake of comparison, the gas-to-dust size ratio of the disk around Oph~IRS63 is slightly larger than the average value obtained in the more evolved T~Tauri stars of $2.9\pm1.2$ \citep{Long2022} and to other values found in the literature \citep[e.g.,][]{Ansdell2018,Trapman2020}. 

Using thermochemical models, \cite{Trapman2019} and \cite{Sanchis2021} found that a gas-to-dust size ratio of $\rm R_{gas}/R_{dust}>4$ are clear sign of dust evolution and radial drift in disks. They also showed that in cases where $\rm R_{gas}/R_{dust}<4$, optical depth effects alone could reproduce these observational signatures. In the latter case, identifying dust evolution from $\rm R_{gas}/R_{dust}$ requires detail modelling of the disk, including the total CO content. For the case of Oph~IRS63, the gas-to-dust size ratio is slightly above the critical value of 4. Thus, it is possible that radial drift effects alone could account for the size difference. However, as mentioned above, the results obtained for Oph~IRS63 might not be directly comparable to studies in Class II sources. Future studies with detail modeling of the dust and gas around Class~I sources will be necessary to clarify whether the radial size difference observed for Oph~IRS63 implies dust growth and radial drift.

\subsection{System structure}
\label{subsec:schematic_view_system}
\begin{figure}[ht!]
\centering
\epsscale{1.1}
\plotone{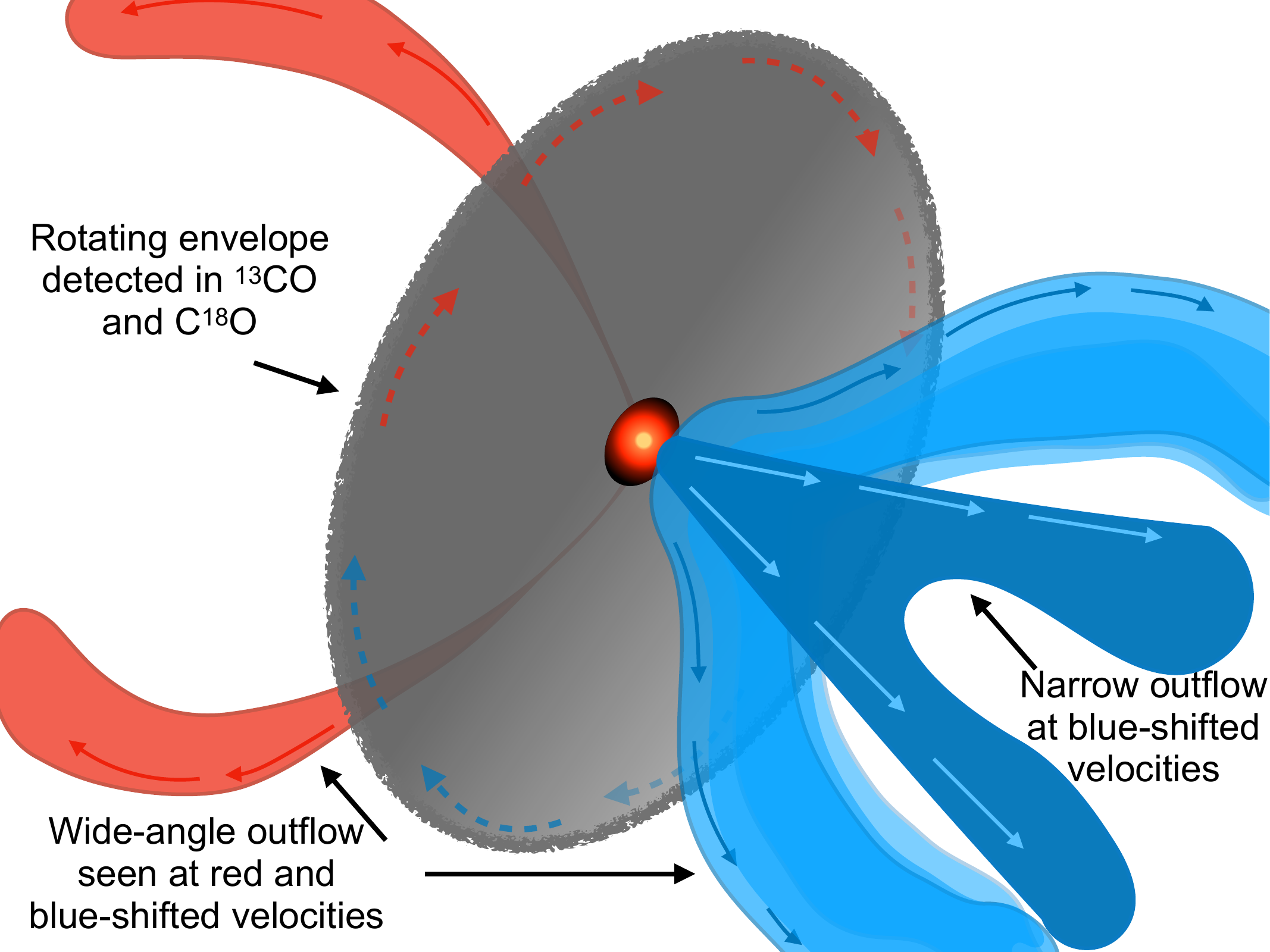}
\caption{Schematic view of the system's large-scale emission. The gray rotating structure represents the envelope observed in $^{13}$CO and C$^{18}$O at radial offset $\geq 2\arcsec$. The red arrows mark the portion of the envelope receding, while the blue one mark emission towards us. The wide outflows and the narrow blue-shifted outflow are also depicted. The blue-shifted outflow partially blocks the view toward the bottom-right part of the envelope due to geometrical effects.}
\label{fig:Cartoon_large-scale}
\end{figure}

Figure \ref{fig:Cartoon_large-scale} shows a schematic view of the large-scale emission of the system. The envelope is represented by the gray rotating flattened structure observed in $^{13}$CO at radial offset $\geq 2\arcsec$. The red arrows indicate the portion of the envelope receding, while the blue one mark emission towards us. The envelope should be at least as large as the FOV of our observations ($\sim25\arcsec$) based on the streamer structure seen towards the northeast in C$^{18}$O, the extended emission seen at low intensity in $^{13}$CO, and the extended emission observed close to the systemic velocity of $^{12}$CO (more specifically in channel maps with velocities of 0.96, 1.6, 3.5, and 4.14 km s$^{-1}$, see Figure \ref{fig:channel_maps_12CO} of Appendix \ref{appendix:channel_maps}).

The almost Keplerian rotating disk structure seen in the observations with angular scale $\leq 2\arcsec$ is represented in colors with an oval shape at the center of the image. Assuming the orientation of the envelope is similar to the disk ($i_{disk}=47^{\circ}$ and P.A. =$149^{\circ}$), and given the outflow directions, we expect the southwest part of the envelope to lie behind the blue-shifted outflow, while the northeast part should be in front (from our perspective) of the red-shifted outflow. This geometry could explain the lack of envelope emission towards the west and southwest seen in the $^{13}$CO peak intensity and moment 0 map of Figure \ref{fig:13CO_map}. Additionally, this geometry partially explains why the red-shifted outflow is weaker and has a simpler structure than the blue-shifted one due to self-absorption and emission being more resolved out. The figure also depicts the blue and red-shifted wide-angle outflows, along with the narrow blue-shifted outflow emission. 


\section{Summary} \label{sec:Summary}

We present the ALMA $^{12}$CO, $^{13}$CO, C$^{18}$O, SO, and H$_2$CO observations of the Class I protostellar source Oph~IRS63 obtained in the context of the eDisk collaboration. Ring-like structures have previously been detected in the dust continuum around Oph IRS63, making it an ideal target to study its gas structure and kinematics. Our findings can be summarized as follows:

\begin{enumerate}
    \item Our ALMA observations reveal a myriad of protostellar features, including an extended rotating envelope in $^{13}$CO, a large-scale arc structure in C$^{18}$O, several small-scale spirals in SO, a quasi-Keplerian disk identified in $^{13}$CO, C$^{18}$O, and SO, and a shell-like bipolar outflow prominent in $^{12}$CO. 

    \item Using the SLAM tool, we find that at radial distance $\rm R >265$~au, the $^{13}$CO emission follows a velocity profile consistent with infalling motion. Similarly, SO follows an infalling motion profile at radial distances greater than $\rm R >110$~au. The moment 1 map of C$^{18}$O and SO provides additional support for the infall profile, as we observe an S-shaped pattern in the zero radial velocity line.
    
    \item At radial distances of less than $\sim 2\arcsec$, we observe a velocity gradient along the major axis of the disk in $^{13}$CO and C$^{18}$O. By fitting a  power law function to the rotational velocity profiles of the lines, we confirm a (quasi)Keplerian rotational profile with $v\propto r^{-0.61\pm0.04}$ out to a distance of $R_{gas}=265 \pm 9$~au. At distances smaller than $\sim$110~au, the SO also shows a Keplerian-like rotation. In both cases, our calculations indicate that these rotational profile are produced by a central source with a protostellar mass of $\rm M_\star = 0.5\pm 0.2 \; M_\odot$. 
        
    \item We successfully reproduced three C$^{18}$O arc-like features as streamers resulting from an infalling rotating envelope using the formalism described by \cite{Mendoza2009}. Additionally, we measured the mass infall rate of these streamers (from envelope to disk) and found a value of $\dot{M}\sim 4 \times 10^{-6}$~$\rm M_\odot \,  yr^{-1}$. This value is much larger than the (proto)stellar mass accretion rate from disk to star of $\dot{M}\sim 2-20 \times 10^{-8}$~$\rm M_\odot \,  yr^{-1}$ we measured from spectroscopy. The natural consequence of this mass transfer imbalance is that the protostellar disk is in a mass build-up phase. If such mass accretion imbalance continues for another $\sim10^4$~yrs, the Toomre-Q parameter of the disk will approach the unity, making the disk gravitationally unstable.
    
    \item Using the \cite{Mendoza2009} streamers formalism, we were also able to reproduce the smaller-scale spiral structures observed in SO. Observationally, the SO spirals connect to the disk's inner part, forming a ring-like structure at a radius of $\sim 0\farcs5\pm 0\farcs2$. Based on the shape and location of the SO ring, we propose that it appears at the interface between the infalling material and the disk, possibly via slow shocks, as suggested by the models proposed by \cite{Aota2015}.
    
    \item We have detected a bipolar outflow in the $^{12}$CO line. The blue-shifted side of the outflow exhibits both a wide- and narrow-angle component, each displaying a complex shell-like structure. The northern part of the red-shifted outflow appears to be spatially close to the C$^{18}$O streamer. We speculate that when the outflow opened the cavity observed in $^{12}$CO, it pushed the envelope creating a localized higher density region that we observe as a C$^{18}$O streamer. The northern streamer might then infall into the star following the rim of the outflow cavity.
    
    \item The gaseous disk measured from $^{13}$CO $R_{gas} = 265 \pm 9$~au is about four times larger than the dust disk $R^{dust}_{95\%}$=62~au. Thermochemical models suggest that such a large difference is a clear signs of dust evolution and drift in Class II disks. However, due to the embedded nature of source, our measurement of the gas disk radius is not completely comparable with these models. Detail modelling of gas and dust in Class I sources would be necessary to clarify the radial size difference observed for Oph~IRS63.
    
\end{enumerate}

Overall, these observations demonstrate the dynamic and rich environment surrounding Oph IRS63. Intricate connections exist between the outflow, infalling envelope, and disk. Understanding mass transfer in the system can help us measure the available material for planet formation during the protostellar phase. Unlike in Class II sources, the mass budget of protostellar disks depends not only on stellar accretion, but also on the envelope feeding it, and to a lesser extent on the material ejected by the outflow.



\newpage
\section*{Acknowledgments}

We appreciate the referee for providing an insightful review that has helped to improve this paper. We are thankful to Michael S. Connelley for helping with determinations of the extinction towards Oph~IRS63, Jesus Lopez-Vazquez for the discussion about the streamer models, and Hsien Shang for helpful comments on the outflow structure. C.F. and N.O. acknowledge support from National Science and Technology Council (NSTC) in Taiwan through grants NSTC 111-2124-M-001-005, NSTC 112-2124-M-001-014, NSTC 109-2112-M-001-051 and 110-2112-M-001-031. S.T. is supported by JSPS KAKENHI Grant Numbers 21H00048 and 21H04495. This work was supported by NAOJ ALMA Scientific Research Grant Code 2022-20A. J.J.T. acknowledges support from NASA XRP 80NSSC22K1159.
J.E.L. was supported by the National Research Foundation of Korea (NRF) grant funded by the Korean government (MSIT) (grant number 2021R1A2C1011718).
J.K.J. and R.S. acknowledge support from the Independent Research Fund Denmark (grant No. 0135-00123B).
Z.-Y.D.L. acknowledges support from NASA 80NSSC18K1095, the Jefferson Scholars Foundation, the NRAO ALMA Student Observing Support (SOS) SOSPA8-003, the Achievements Rewards for College Scientists (ARCS) Foundation Washington Chapter, the Virginia Space Grant Consortium (VSGC), and UVA research computing (RIVANNA).
Z.Y.L. is supported in part by NASA 80NSSC20K0533 and NSF 2307199 and 1910106.
I.d.G.-M. acknowledges support from grant PID2020-114461GB-I00, funded by MCIN/AEI/10.13039/501100011033.
P.M.K. acknowledges support from NSTC 108-2112- M-001-012, NSTC 109-2112-M-001-022 and NSTC 110-2112-M-001-057.
W.K. was supported by the National Research Foundation of Korea (NRF) grant funded by the Korea government (MSIT) (NRF-2021R1F1A1061794). S.P.L. and T.J.T. acknowledge grants from the NSTC of Taiwan 106-2119-M-007-021-MY3 and 109-2112-M-007-010-MY3.
C.W.L. is supported by the Basic Science Research Program through the NRF funded by the Ministry of Education, Science and Technology (NRF- 2019R1A2C1010851), and by the Korea Astronomy and Space Science Institute grant funded by the Korea government (MSIT; Project No. 2022-1-840-05). 
L.W.L. acknowledges support from NSF AST-2108794.
S.N. acknowledges support from the National Science Foundation through the Graduate Research Fellowship Program under Grant No. 2236415. 


J.P.W. acknowledges support from NSF AST-2107841.
Y.A. acknowledges support by NAOJ ALMA Scientific Research Grant code 2019-13B, Grant-in-Aid for Scientific Research (S) 18H05222, and Grant-in-Aid for Transformative Research Areas (A) 20H05844 and 20H05847.
M.L.R.H. acknowledges support from the Michigan Society of Fellows.
H.-W.Y. acknowledges support from the NSTC in Taiwan through the grant NSTC 110-2628-M-001-003-MY3 and from the Academia Sinica Career Development Award (AS-CDA-111-M03).
This paper makes use of the following ALMA data: ADS/JAO.ALMA$\#$2019.A.00034.S and ADS/JAO.ALMA$\#$2015.1.01512.S. ALMA is a partnership of ESO (representing its member states), NSF (USA) and NINS (Japan), together with NRC (Canada), MOST and ASIAA (Taiwan), and KASI (Republic of Korea), in cooperation with the Republic of Chile. The Joint ALMA Observatory is operated by ESO, AUI/NRAO and NAOJ.
The National Radio Astronomy Observatory is a facility of the National Science Foundation operated
under cooperative agreement by Associated Universities, Inc.

%

\vspace{5mm}
\facilities{ALMA, IRTF}


\software{CASA \citep{McMullin2007}, SLAM \citep{SLAM}, GoFish \citep{GoFish}, SpexTool \citep{Cushing2004}, CARTA \citep{CARTA}}



\appendix
\restartappendixnumbering

\section{Infrared spectra of Oph~IRS63}
\label{appendix:infrared}

To calculate the protostellar mass accretion rate, we used spectroscopic observations carried out with the 3.2m NASA Infrared Telescope Facility (IRTF) on Maunakea, Hawaii. Oph~IRS63 was observed with iSHELL \citep{Rayner2022} on 2020 August 23 in the K-band (K2 mode) with the $0\farcs75$ slit-width configuration (Flores et al. submitted). We also use SpeX observations in the SXD mode \citep{Cushing2004,Vacca2003}, i.e., from 0.7~$\mu m$ to 2.5~$\mu m$ with the $0\farcs5$ slit width $R\sim1200$ on 2020 August 19. 


We measured the optical extinction to Oph~IRS63 in two ways, first using the column densities derived from ro-vibrational CO lines redward of 2.35~$\mu$m in the iSHELL spectrum. Flores et al. (submitted) measured $^{12}$CO column densities of $2 - 5 \times$ 10$^{18}$ cm$^{-2}$, which assuming a standard abundance of $^{12}$CO of 10$^5$ correspond to H$_2$ column densities of $2-5 \times$ 10$^{22}$ cm$^{-2}$ or equivalently optical magnitudes of $A_v \sim 20 - 50$ mag.

Secondly, using the method described in \cite{Connelley2010}, where the spectrum of Oph~IRS63 is fit using a black body component (with a fixed veiling of $\rm r_{k}=3$, based on iSHELL observations) and the reddened spectrum of a star with a similar spectral type to Oph~IRS63 (M2 or $\rm T_{eff}\sim3500$~K) using the extinction law from \cite{Fitzpatrick1999}. The second procedure result in extinction estimates between 20 to 40 magnitudes of visual extinction with a best fit of 24 magnitudes, in agreement with the extinction predicted from iSHELL data. To de-redden the iSHELL spectrum and measure the Br$\gamma$ line luminosity, we adopt 24 magnitudes of extinction as our best estimate.

\section{Channel maps}
\label{appendix:channel_maps}

In the figures below, we show the channel maps of the $^{12}$CO, $^{13}$CO, C$^{18}$O, SO, and H$_2$CO lines. Detailed information of the molecules is presented in Table \ref{tab:observation_info}. The color scale is stretched using the inverse hyperbolic sine function, we use blue contours to mark negative emission, and in each channel map, we mark the protostellar position with a yellow cross.

\begin{figure*}[ht!]
\centering
\epsscale{1.1}
\plotone{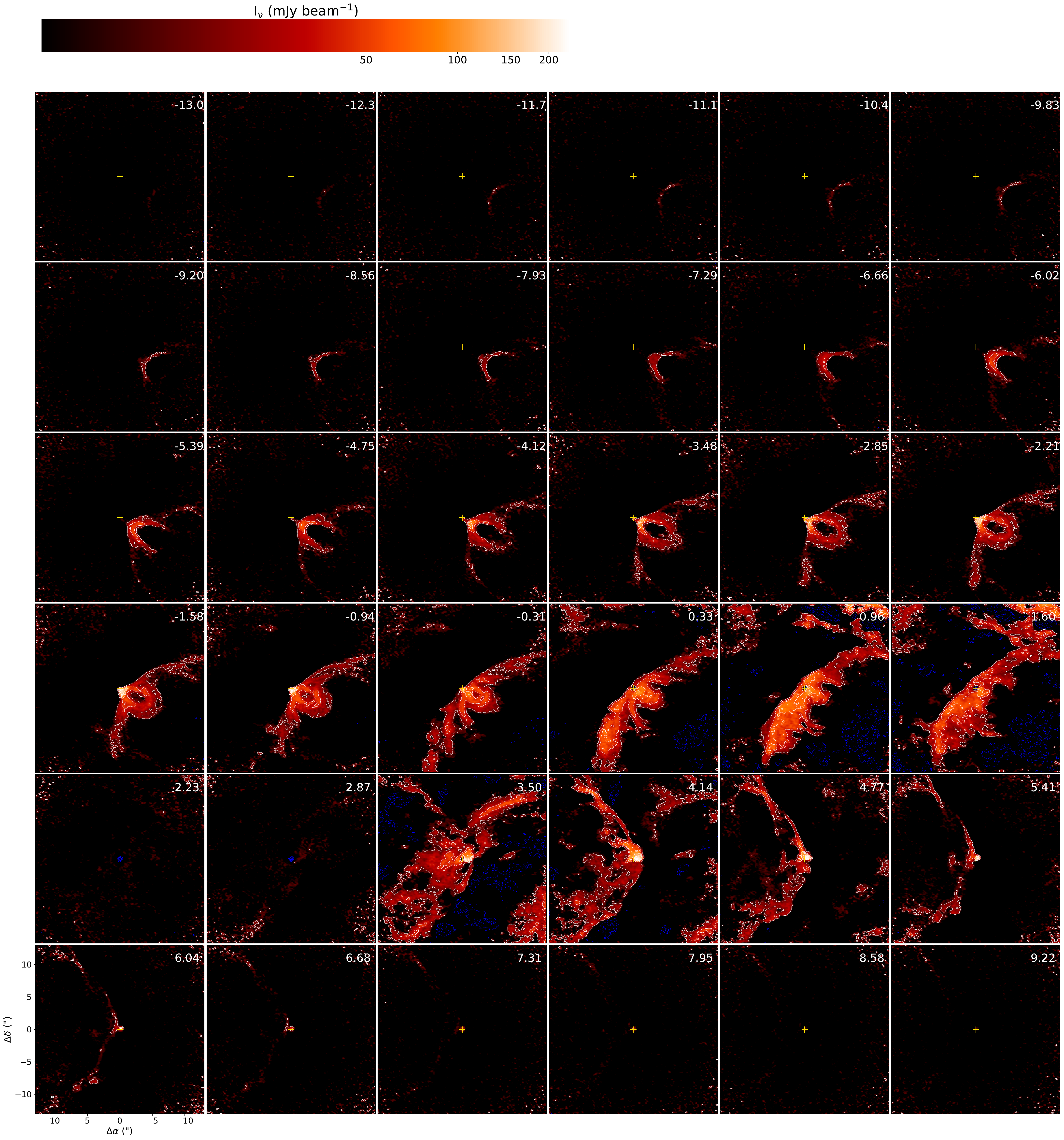}
\caption{Channel map of the $^{12}$CO emission in Oph IRS63. The lowest contour emission corresponds to 6$\sigma$ or 12 \mjbeam, with increments of 12$\sigma$ at a time. The FOV of the map is $26 \arcsec \times26 \arcsec$.}
\label{fig:channel_maps_12CO}
\end{figure*}

\begin{figure*}[ht!]
\centering
\epsscale{1.1}
\plotone{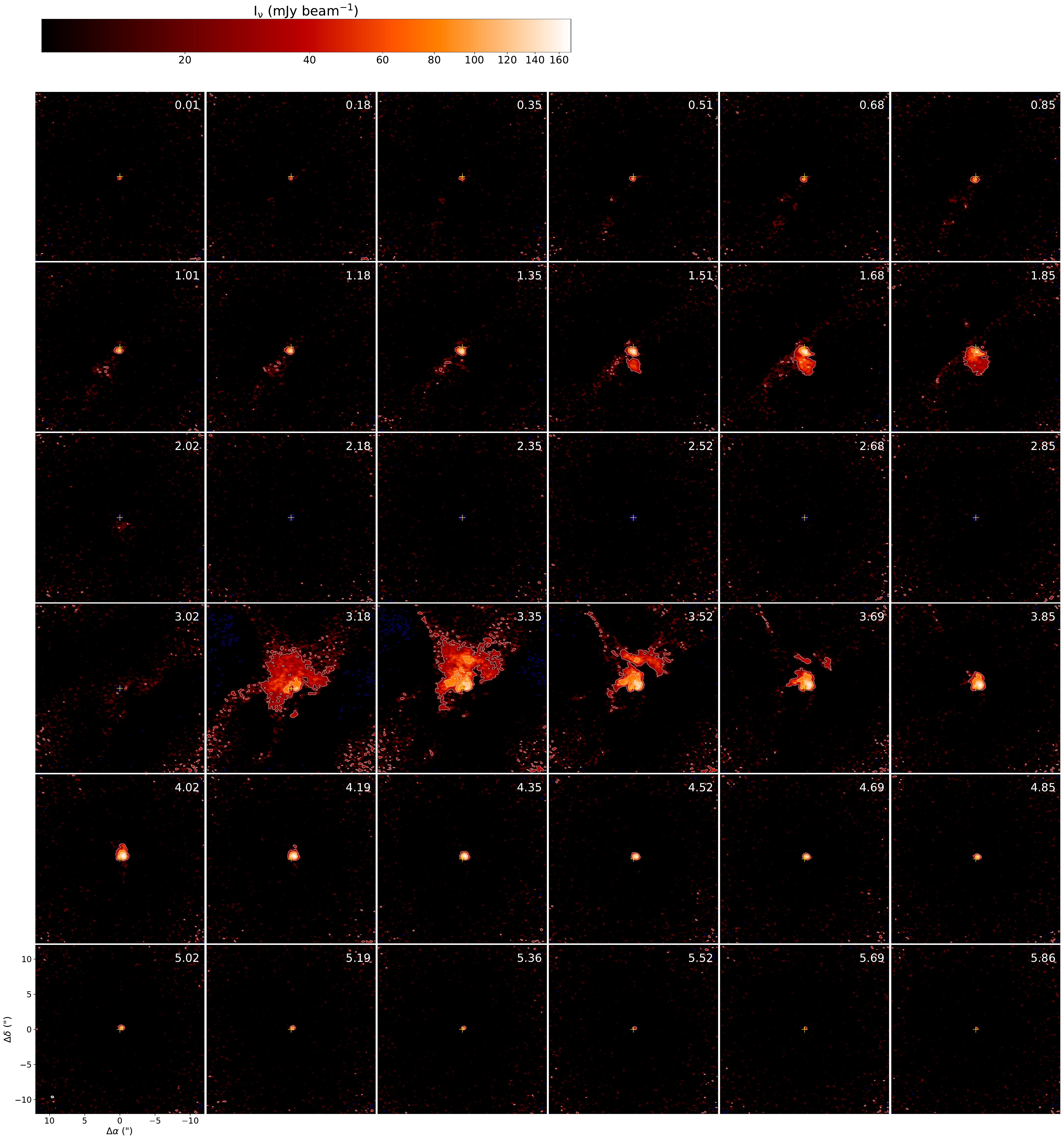}
\caption{Channel map of the $^{13}$CO emission in Oph IRS63. The lowest contour emission corresponds to 5$\sigma$ or 22.5 \mjbeam, with increments of 10$\sigma$ at a time. The FOV of the map is $24 \arcsec\times24 \arcsec$.}
\label{fig:channel_maps_13CO}
\end{figure*}

\begin{figure*}[ht!]
\centering
\epsscale{1.1}
\plotone{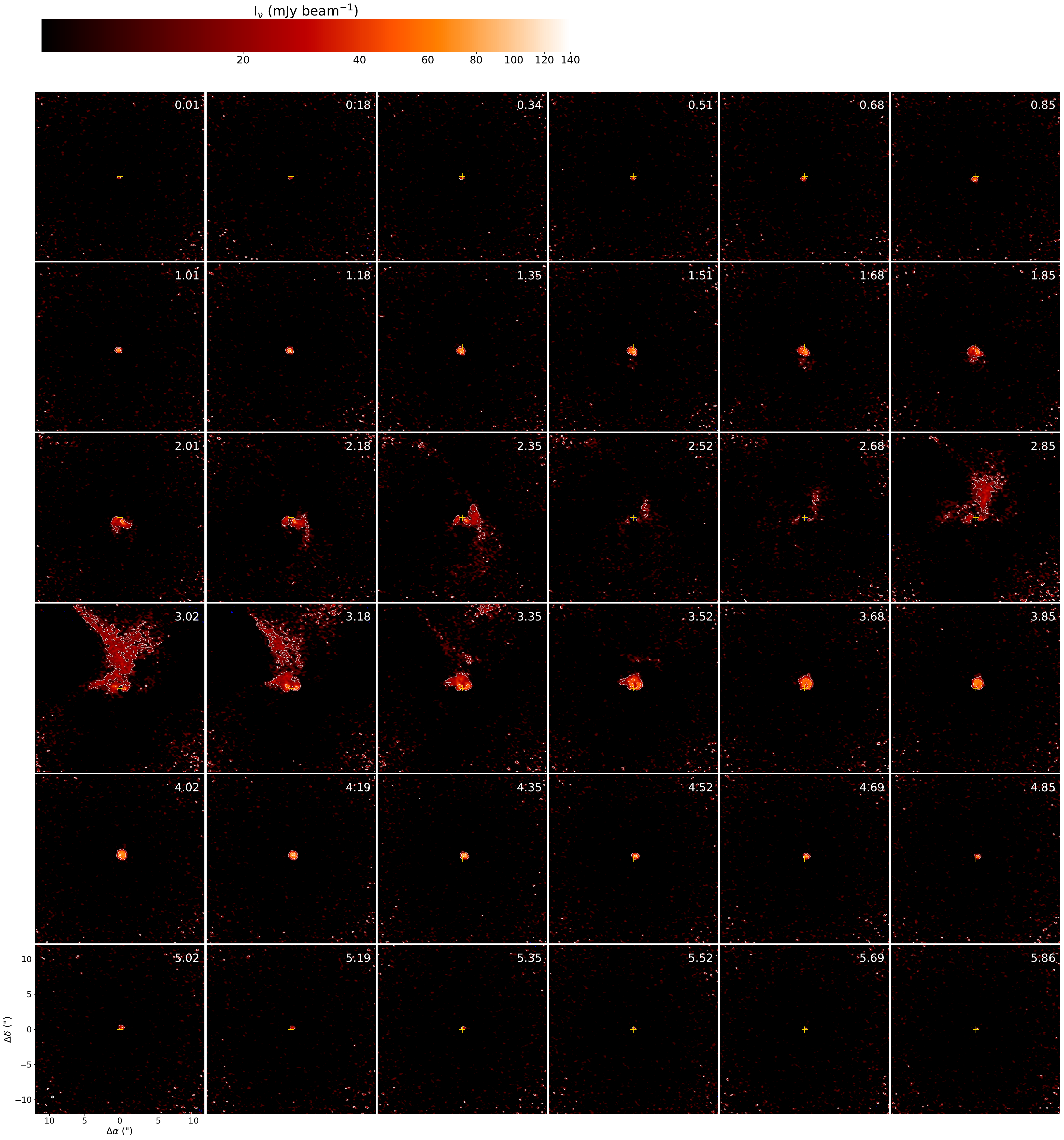}
\caption{Channel map of the C$^{18}$O emission in Oph IRS63. The lowest contour emission corresponds to 5$\sigma$ or 15 \mjbeam, with increments of 10$\sigma$ at a time. The FOV of the map is $24 \arcsec \times24 \arcsec$.}
\label{fig:channel_maps_C18O_large}
\end{figure*}


\begin{figure*}[ht!]
\centering
\epsscale{1.1}
\plotone{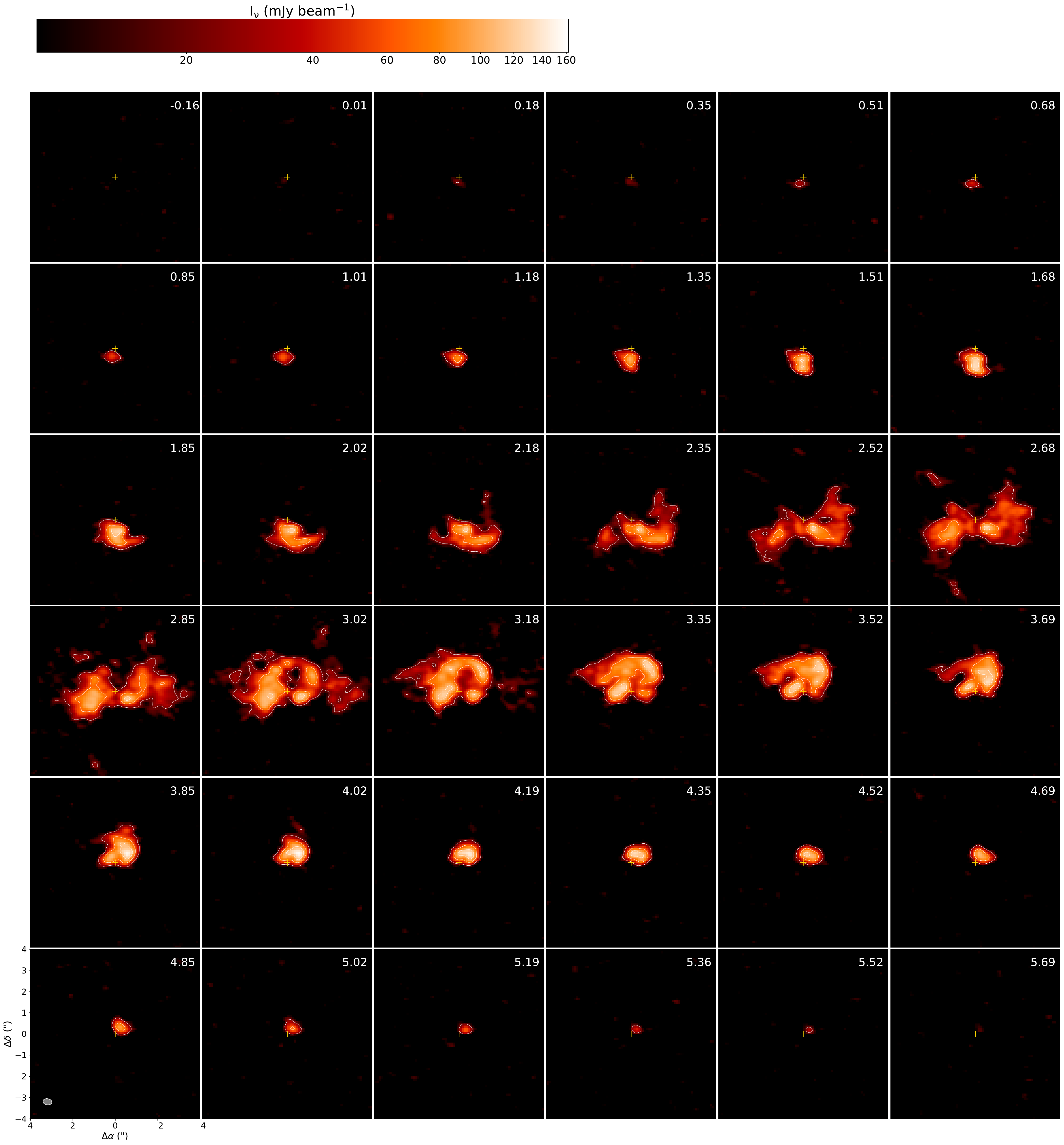}
\caption{Channel map of the SO emission in Oph IRS63. The lowest contour emission corresponds to 5$\sigma$ or 22 \mjbeam, with increments of 10$\sigma$ at a time. The FOV of the map is $8 \arcsec \times8 \arcsec$.}
\label{fig:channel_maps_SO}
\end{figure*}

\begin{figure*}[ht!]
\centering
\epsscale{0.7}
\plotone{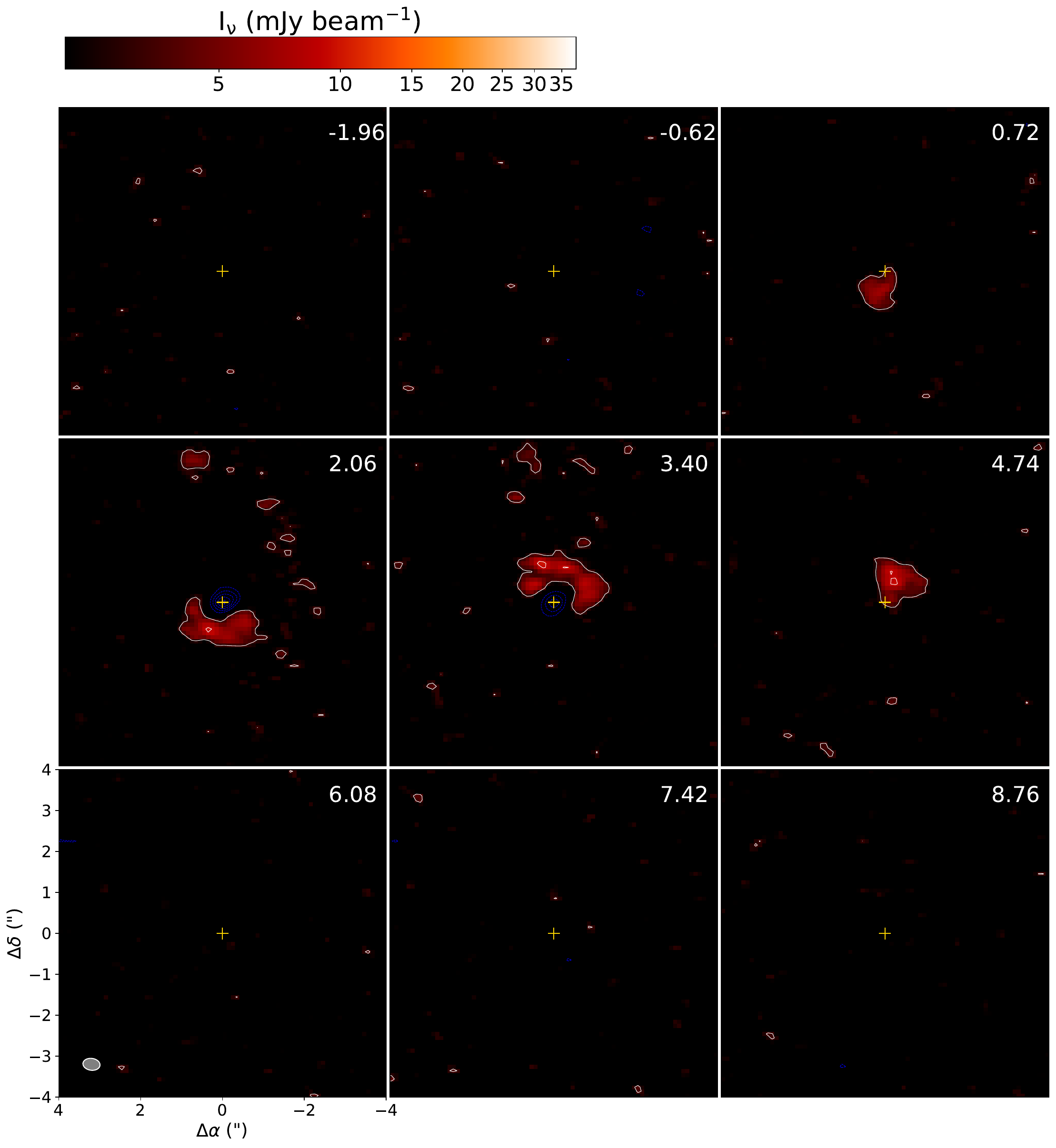}
\caption{Channel map of formaldehyde (H$_2$CO) emission in Oph IRS63. The lowest contour emission corresponds to 3$\sigma$ or 3 \mjbeam, with increments of 3$\sigma$ at a time. The FOV of the map is $8\arcsec \times8 \arcsec$.}
\label{fig:channel_maps_H2CO}
\end{figure*}


\bibliography{sample631}{}
\bibliographystyle{aasjournal}



\end{document}